\documentclass[aps,prd,showpacs,twocolumn]{revtex4}

\usepackage{epsfig}
\usepackage{longtable}
\usepackage{morefloats}
\usepackage{bm}

\begin{document}

\def\nn{$n\bar n$ }
\def\qq{$q\bar q$ }
\def\cc{$c\bar c$ }
\def\ccbar{$c\bar c$ }

\def\x{\bf x}
\def\B{\rm B}
\def\D{\rm D}
\def\E{\rm E}
\def\F{\rm F}
\def\G{\rm G}
\def\H{\rm H}
\def\I{\rm I}
\def\J{\rm J}
\def\K{\rm K}
\def\L{\rm L}
\def\M{\rm M}
\def\P{\rm P}
\def\S{\rm S}
\def\T{\rm T}
\def\V{\rm V}

\title{$B$ and $B_s$ Meson Spectroscopy} 
\author{
Stephen Godfrey$^a$\footnote{Email: godfrey@physics.carleton.ca},
Kenneth Moats$^a$\footnote{Email: kmoats@physics.carleton.ca}
and
Eric S. Swanson$^b$\footnote{Email: swansone@pitt.edu}
}
\affiliation{
$^a$Ottawa-Carleton Institute for Physics, 
Department of Physics, Carleton University, 
Ottawa K1S 5B6, Canada\\
$^b$Department of Physics and Astronomy, University of Pittsburgh, 
Pittsburgh PA 15260.}

\date{August 29, 2016}

\begin{abstract}
Properties of bottom and bottom-strange mesons are computed in two relativized quark models. Model masses and wavefunctions are used to predict radiative transition rates and the $^3P_0$ quark pair creation model is used to compute strong decay widths. A comparison to recently observed bottom and bottom-strange states is made. 
We find that there are numerous excited $B$ and $B_s$ 
mesons that have relatively narrow widths and significant branching ratios to simple final states such
as $B\pi$, $B^*\pi$, $BK$, and $B^*K$ that could be observed in the near future.
\end{abstract}

\maketitle

\section{Introduction}

Meson spectroscopy has undergone a renaissance in recent years with the discovery of many new hadronic states,
both those that are well described by quark models but also  many poorly understood states
such as the enigmatic $XYZ$ charmonium and bottomonium-like states \cite{Godfrey:2008nc,Eichten:2007qx,Godfrey:2009qe}. 
In the bottom meson sector, the CDF and D0 collaborations at Fermilab, and more recently the 
LHCb experiment at CERN \cite{Aaij:2015qla}, 
have observed  $P$-wave $B$ and $B_s$ states.
In parallel to these advancements in experiment, lattice QCD is also making strides in calculating
hadron masses and other properties \cite{Lang:2015hza}.  Progress in both experiment
and theory go hand in hand in advancing our understanding of hadron physics and QCD in the 
soft regime.  Understanding the properties of the $B$ mesons can play an important role in
this enterprise as, in the heavy quark limit, $B$ mesons can be viewed as 
the hydrogen ``atoms'' of QCD with a light quark
interacting with a heavy static quark.  Understanding the $B$ mesons will give 
a more complete understanding of excited mesons and will also help put the newly 
discovered excited charmed mesons into the larger context.  The significantly higher statistics expected at the LHC
increases the likelihood of observing many new excited $B$ mesons which will give us
the opportunity to study $B_{(s)}$ meson spectroscopy in greater detail than previously possible. 
In anticipation of these experimental developments 
it will be useful to predict the properties of these states,
both as guidance to help experimental searches, and also to test our theoretical understanding against
experimental measurements once they have been observed.  

In this paper we study $B$ meson spectroscopy using the constituent quark model to calculate masses
and wavefunctions.  For unequal mass quarks and antiquarks $C$-parity is not a good quantum
number so that states with the same parity and spin can mix (such as the $^3P_1$ and $^1P_1$
states).  The relevant mixing angle is also calculated using the quark model.  
The wavefunctions are used to calculate radiative transition widths and
as input to calculate strong decays using the $^3P_0$ quark pair creation model.  We use
two different relativistic quark models to gauge variations in predictions from details of the
models.  The quark models are based on a relativistic kinetic energy term with a 
short distance one-gluon-exchange potential with a strong coupling constant that runs and a 
linear confining potential 
\cite{Godfrey:1985xj,Godfrey:1986wj,Godfrey:2004ya,Godfrey:2005ww}.  
We make predictions of 
properties which will be useful for both finding and understanding these states which are the 
mass predictions,  E1 and M1 radiative
transitions, and strong partial and total decay widths obtained using the $^3P_0$ pair creation model for states
above threshold.  
We put these results together 
to help identify recently observed excited $B$ mesons and 
to discuss the most likely means of observing excited $B$ and $B_s$ mesons and 
strategies for searching for them.

\section{Spectroscopy}

We consider two relativized quark models.  The first is the model of Godfrey and Isgur \cite{Godfrey:1985xj}
which has been a useful guide to mesons, from the lightest isovector mesons to the heaviest
bottomonium states.  The second model is due to Swanson and collaborators 
\cite{LS,ls2,ess} which incorporates recent developments in effective field theory and lattice
gauge theory.  By considering two models we may be able to better gauge the predictive 
limitations of our results.

\subsection{The relativized quark model}
\label{sec:rqm}

The relativized quark model incorporates
a relativistic dispersion relation for the quark kinetic energy and an instantaneous interaction  
comprised of a short distance one-gluon-exchange Lorentz vector potential and a Lorentz scalar
linear confining potential:
\begin{equation}
{\rm H} = {\rm H}_0 + \V_{q\bar{q}} (\vec{p},\vec{r})
\end{equation}
where the relativistic kinetic term is given by 
\begin{equation}
{\rm H}_0 = 
\sqrt{\vec p_q^{\, 2} +m_q^2} 
+ \sqrt{\vec p_{\bar q}^{\, 2} + m_{\bar{q}}^2 }\ .
\end{equation}
Just as in the nonrelativistic model, the quark-antiquark potential 
$\V_{q\bar{q}} (\vec{p},\vec{r})$ assumed here incorporates spin-dependent interactions that arise from the nonrelativistic reduction of the full interaction.
The colour Coulomb potential and the spin dependent 
potentials arising from one-gluon-exchange include
a QCD-motivated running coupling $\alpha_s(r)$, all terms in the potential are modified by a flavor-dependent 
potential smearing parameter $\sigma$, and  
quark masses in the spin-dependent interactions are replaced with quark kinetic energies.
To first order in $(v_q/c)^2$, $\V_{q\bar{q}} (\vec{p},\vec{r}\,)$
reduces to the standard non-relativistic result.
Details of the model and the method of solution can be found in Ref. \cite{Godfrey:1985xj}. 

For the case of a quark and antiquark of unequal mass charge conjugation
parity is no longer a good quantum number so that states with different 
total spins but with the same total angular momentum, such as
the $^3P_1 -^1P_1$ and $^3D_2 -^1D_2$ pairs, can mix via
the spin orbit interaction or some other mechanism.
Consequently, the physical $j=1$ $P$-wave states are linear
combinations of $^3P_1$ and $^1P_1$ which we describe by:
\begin{eqnarray}
\label{eqn:mixing}
P  & = & {}^1P_1 \cos\theta_{nP} + {}^3P_1 \sin\theta_{nP} \nonumber \\   
P' & = & -{}^1P_1 \sin\theta_{nP} + {}^3P_1 \cos \theta_{nP} 
\end{eqnarray}
with analogous notation for the corresponding $L=D$, $F$, etc. pairs.
In Eq. \ref{eqn:mixing},
$P\equiv L=1$ designates the relative angular momentum of the $q\bar{q}$ 
pair and the subscript $J=1$ is the total angular momentum of the $q\bar{q}$ 
pair which is equal to $L$.  Our notation implicitly implies $L-S$ 
coupling between the quark spins and the relative orbital angular momentum.  
In the heavy quark limit in which 
the heavy quark mass $m_Q\to \infty$, 
the states can be described by the total angular momentum of the
light quark, $j$, which couples to the spin of the heavy quark and
corresponds to $j-j$ coupling.  This limit gives
rise to two doublets which for $L=1$ have $j=1/2$ and  
$j=3/2$ and corresponds to two physically independent mixing angles 
$\theta_{1P}=-\tan^{-1}(\sqrt{2})\simeq -54.7^\circ$ and 
$\theta_{1P}=\tan^{-1}(1/\sqrt{2})\simeq 35.3^\circ$ \cite{barnes}. Some 
authors prefer to use the $j-j$ basis \cite{eichten94}
but since we solve our 
Hamiltonian equations assuming $L-S$ eigenstates and then include the 
$LS$ mixing we use the notation of Eq. \ref{eqn:mixing}.  
It is straightforward to
transform between the $L-S$ basis and the $j-j$ basis.  It will turn 
out that radiative transitions are particularly sensitive to the 
$^3L_L-^1L_L$ mixing angle with predictions from different 
models in some cases
giving radically different results.  We also note that the 
definition of the mixing angles are fraught with ambiguities.  For 
example, charge conjugating $q\bar{b}$ into $b\bar{q}$ flips the 
sign of the angle and the phase convention depends on the
order of coupling $\vec{L}$, $\vec{S}_q$, and $\vec{S}_{\bar{q}}$
\cite{barnes}.

\begin{figure*}[t]
\includegraphics[width=14cm,angle=0]{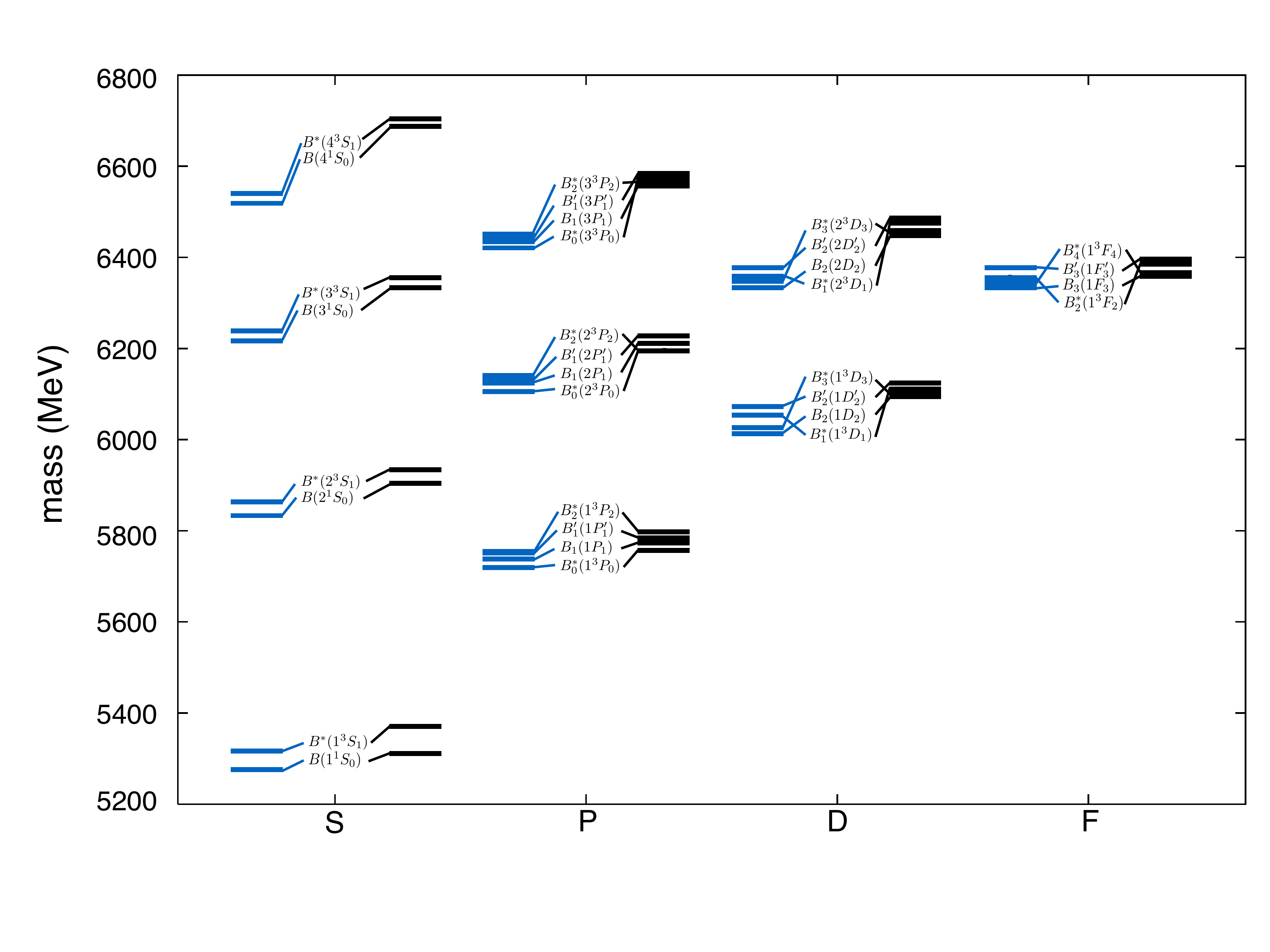}
\caption{Bottom mass predictions for the 
AR model (blue and left set of lines) and Godfrey-Isgur model
(black and right set of lines).}
\label{fig:1}
\end{figure*}

\begin{figure*}[t]
\includegraphics[width=14cm,angle=0]{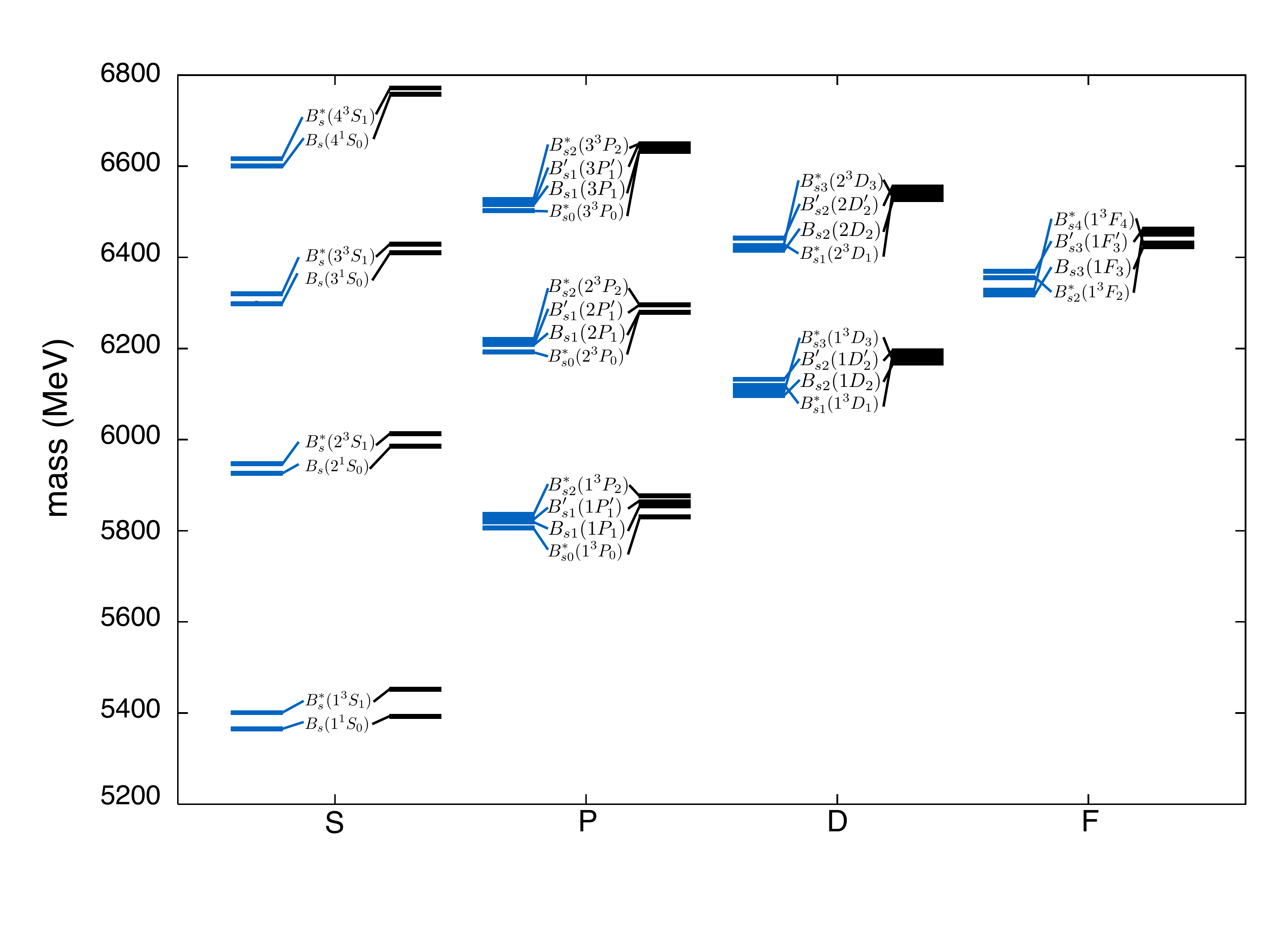}
\caption{Strange bottom mass predictions for the 
AR model (blue and left set of lines) and Godfrey-Isgur model
(black and right set of lines).}
\label{fig:2}
\end{figure*}

The  Hamiltonian problem was solved 
using the following parameters: the slope of the
linear confining potential is 0.18 GeV$^2$, 
$m_q=0.220$~GeV, $m_s=0.419$~GeV and $m_b=4.977$~GeV.  Other parameters can be found in Ref. \cite{Godfrey:1985xj}.  Predicted masses and mixing angles 
are given in Figs.~\ref{fig:1} and \ref{fig:2} and in 
Tables \ref{tab:masses}-\ref{tab:masses2}.

\begin{table*}
\caption{Predicted masses (in MeV), spin-orbit mixing angles and effective harmonic oscillator parameters, $\beta_{eff}$ (in GeV).  
Columns 2-5 show the results using the Godfrey-Isgur relativized quark model described in Sec. \ref{sec:rqm}
and columns 6-9 show the results using the alternate relativized model described in Sec. \ref{sec:arm}.  The $P_1-P_1'$, $D_2-D_2'$, 
$F_3-F_3'$ and $G_4-G_4'$ 
states and mixing angles
are defined using the convention of Eq. \ref{eqn:mixing}.  Where two values of $\beta_{eff}$ are listed, the first (second) refers to the singlet (triplet) state.
\label{tab:masses}}
\begin{tabular}{l l c l c l c l c } \hline \hline
  & \multicolumn{2}{c}{GI $b\bar{q}$}        & \multicolumn{2}{c}{GI $b\bar{s}$}  & \multicolumn{2}{c}{ARM $b\bar q$}   & \multicolumn{2}{c}{ARM $b\bar s$} \\ 
  State                  & Mass  & $\beta_{eff}$  & Mass  & $\beta_{eff}$   & Mass  & $\beta_{eff}$  & Mass  & $\beta_{eff}$ \\ 
\hline\hline
$1^3S_1 $       	& 5371  & 0.542                         & 5450  & 0.595     & 5316 & 0.586  & 5400 & 0.616 \\
$1^1S_0 $       	& 5312  & 0.580                         & 5394  & 0.636     & 5275 & 0.628  & 5366 & 0.651 \\
\hline
$1^3P_2 $    	& 5797  & 0.472                         & 5876  & 0.504     & 5754 & 0.465  & 5836 & 0.487  \\
$1 P_1$         	& 5777  & 0.499, 0.511           	& 5857  & 0.528, 0.538  & 5738 & 0.481, 0.492 & 5822 & 0.500, 0.507 \\
$1 P_1'$        	& 5784  &                       & 5861  &                   & 5753 &  & 5830 & \\
$1^3P_0 $       	& 5756  & 0.536                         & 5831  & 0.563     & 5720 & 0.525 & 5805 & 0.531  \\
$\theta_{1P}$   & \multicolumn{2}{l}{30.28$^\circ$}     & \multicolumn{2}{l}{39.12$^\circ$}    & \multicolumn{2}{l}{43.6$^\circ$}  & \multicolumn{2}{l}{37.9$^\circ$} \\
\hline
$2^3S_1 $       & 5933  & 0.468                         & 6012  & 0.497   & 5864 & 0.460 & 5948 & 0.477  \\
$2^1S_0 $       & 5904  & 0.477                         & 5984  & 0.508   & 5834 & 0.476 & 5925 & 0.489  \\
\hline
$1^3D_3 $       & 6106  & 0.444                         & 6179  & 0.467  & 6026 & 0.428  & 6109 & 0.443    \\
$1 D_2 $               & 6095  & 0.469, 0.463           & 6169  & 0.482, 0.487  & 6012 & 0.434, 0.436 & 6098 & 0.449, 0.450\\
$1 D_2' $                & 6124  &                      & 6196  &               & 6072 & & 6133 & \\
$1^3D_1 $       & 6110  & 0.488                         & 6182  & 0.504  & 6053 & 0.447  & 6119 & 0.459    \\
$\theta_{1D}$   & \multicolumn{2}{l}{39.69$^\circ$}     & \multicolumn{2}{l}{40.00$^\circ$}  & \multicolumn{2}{l}{48.7$^\circ$} & \multicolumn{2}{l}{48.0$^\circ$} \\
\hline
$2^3P_2 $       & 6213  & 0.440                         & 6295  & 0.462  & 6141 & 0.413 & 6220 & 0.428  \\
$2 P_1 $               & 6197  & 0.452, 0.456           & 6279  & 0.472, 0.474  & 6126 & 0.423, 0.426 & 6208 & 0.435, 0.438\\
$2 P_1' $                & 6228  &                      & 6296  &               & 6132 & & 6211 & \\
$2^3P_0 $       & 6213  & 0.466                         & 6279  & 0.483   & 6106 & 0.439 & 6191 & 0.447    \\
$\theta_{2P}$   & \multicolumn{2}{l}{32.28$^\circ$}     & \multicolumn{2}{l}{33.05$^\circ$}   & \multicolumn{2}{l}{39.35$^\circ$} & \multicolumn{2}{l}{32.1$^\circ$} \\
\hline
$3^3S_1 $       & 6355  & 0.437                         & 6429  & 0.456  & 6240 & 0.412  & 6319 & 0.424  \\
$3^1S_0 $       & 6335  & 0.442                         & 6410  & 0.462  & 6216 & 0.421  & 6301 & 0.431 \\
\hline
$1^3F_4 $       & 6364  & 0.429                         & 6432  & 0.446  & 6244 & 0.408  & 6328 & 0.419     \\
$1 F_3 $               & 6358  & 0.442, 0.446                  & 6425  & 0.457, 0.460  & 6231 & 0.408, 0.408 & 6318& 0.421, 0.421\\
$1 F_3' $                & 6396  &                      & 6462  &          & 6316 & & 6369 & \\
$1^3F_2 $       & 6387  & 0.460                         & 6454  & 0.472    & 6302 & 0.409  & 6358 & 0.423   \\
$\theta_{1F}$   & \multicolumn{2}{l}{41.13$^\circ$}     & \multicolumn{2}{l}{41.14$^\circ$}     & \multicolumn{2}{l}{48.05$^\circ$} & \multicolumn{2}{l}{47.7$^\circ$} \\
\hline
$2^3D_3 $       & 6460  & 0.424                         & 6535  & 0.442   & 6347 & 0.394 & 6428 & 0.405   \\
$2 D_2 $               & 6450  & 0.438, 0.434           & 6526  & 0.449, 0.452  & 6334 & 0.399, 0.401 & 6417 & 0.410, 0.411 \\
$2 D_2' $                & 6486  &                      & 6553  &               & 6377 & & 6442 & \\
$2^3D_1 $       & 6475  & 0.448                         & 6542  & 0.460      & 6357 & 0.410  & 6427 & 0.418  \\
$\theta_{2D}$   & \multicolumn{2}{l}{38.96$^\circ$}     & \multicolumn{2}{l}{39.46$^\circ$}   & \multicolumn{2}{l}{48.2$^\circ$} & \multicolumn{2}{l}{47.3$^\circ$} \\
\hline
$3^3P_2 $       & 6570  & 0.422                         & 6648  & 0.439    & 6451 & 0.388 & 6527 & 0.399     \\
$3 P_1 $               & 6557  & 0.430, 0.432           & 6635  & 0.445, 0.445 & 6438 & 0.394, 0.395 & 6516 & 0.403, 0.404 \\
$3 P_1' $                & 6585  &                      & 6650  &              & 6443 & &  6519 & \\
$3^3P_0 $       & 6576  & 0.437                         & 6639  & 0.451  & 6422 & 0.401 & 6504 & 0.409\\
$\theta_{3P}$   & \multicolumn{2}{l}{31.57$^\circ$}     & \multicolumn{2}{l}{31.59$^\circ$}   & \multicolumn{2}{l}{41.9$^\circ$} & \multicolumn{2}{l}{39.2$^\circ$}  \\
\hline
$1^3G_5 $       	& 6592  & 0.419                         & 6654  & 0.433   & 6433  & 0.395 & 6517  & 0.403     \\
$1 G_4 $          	& 6588  & 0.429 , 0.431             & 6650  & 0.441, 0.443   & 6420  & 0.392, 0.391  & 6507 & 0.403, 0.402\\
$1 G_4' $        	& 6628  &                           & 6690  &                & 6521  &  & 6568 & \\
$1^3G_3 $       	& 6622  & 0.442                         & 6685  & 0.452    & 6508  & 0.389  & 6558  & 0.402   \\
$\theta_{1G}$  	& \multicolumn{2}{l}{41.90$^\circ$}     & \multicolumn{2}{l}{41.87$^\circ$}     & \multicolumn{2}{l}{47.5$^\circ$} & \multicolumn{2}{l}{$47.3^\circ$} \\
\hline \hline
\end{tabular}
\end{table*}

\begin{table*}
\caption{Predicted masses (in MeV), spin-orbit mixing angles and effective harmonic oscillator parameters, $\beta_{eff}$ (in GeV).  
Columns 2-5 show the results using the Godfrey-Isgur relativized quark model described in Sec. \ref{sec:rqm}
and columns 6-9 show the results using the alternate relativized model described in Sec. \ref{sec:arm}.  The $P_1-P_1'$, $D_2-D_2'$, $F_3-F_3'$ and $G_4-G_4'$ states and mixing angles
are defined using the convention of Eq. \ref{eqn:mixing}.  Where two values of $\beta_{eff}$ are listed, the first (second) refers to the singlet (triplet) state.
\label{tab:masses2}}
\begin{tabular}{l l c l c l c l c } \hline \hline
  & \multicolumn{2}{c}{GI $b\bar{q}$}        & \multicolumn{2}{c}{GI $b\bar{s}$}  & \multicolumn{2}{c}{ARM $b\bar q$}   & \multicolumn{2}{c}{ARM $b\bar s$} \\ 
  State                  & Mass  & $\beta_{eff}$  & Mass  & $\beta_{eff}$   & Mass  & $\beta_{eff}$  & Mass  & $\beta_{eff}$ \\ 
\hline\hline
$2^3F_4 $    	& 6679 & 0.414                        & 6748  & 0.428  & 6524  & 0.383  & 6605  & 0.392     \\
$2 F_3 $          	& 6673  & 0.423, 0.425             & 6742  & 0.435, 0.436  & 6511  & 0.384, 0.384  & 6595 & 0.394, 0.392 \\
$2 F_3' $          	& 6711  &                          & 6775  &               & 6583  & & 6638  & \\
$2^3F_2 $       	& 6704  & 0.434                         & 6768  & 0.443    & 6568 & 0.387  & 6627  & 0.397   \\
$\theta_{2F}$   	& \multicolumn{2}{l}{40.86$^\circ$}     & \multicolumn{2}{l}{40.97$^\circ$}     & \multicolumn{2}{l}{47.9$^\circ$} & \multicolumn{2}{l}{47.5$^\circ$} \\
\hline
$4^3S_1 $       & 6703  & 0.421                         & 6773  & 0.436   & 6541 & 0.387  & 6617 & 0.397  \\
$4^1S_0 $       & 6689  & 0.424                         & 6759  & 0.440   & 6520 & 0.393  & 6601 & 0.402  \\
\hline
$3^3D_3 $       	& 6775  & 0.418                         & 6849   & 0.426   & 6623  & 0.375 & 6699  & 0.384   \\
$3 D_2 $         	& 6767  & 0.419, 0.421            & 6841  & 0.431, 0.432   & 6610  & 0.379, 0.381  & 6689 & 0.388, 0.389  \\
$3 D_2' $       	& 6800  &                         & 6864  &                & 6642 & & 6708 & \\
$3^3D_1 $      	& 6792  & 0.427                         & 6855  & 0.438      & 6622 & 0.388  & 6693  & 0.394  \\
$\theta_{3D}$   	& \multicolumn{2}{l}{38.56$^\circ$}     & \multicolumn{2}{l}{39.14$^\circ$}   & \multicolumn{2}{l}{47.7$^\circ$} & \multicolumn{2}{l}{46.8$^\circ$} \\
\hline
$4^3P_2 $       	& 6883  & 0.411                         & 6956  & 0.424    & 6717  & 0.372  & 6790  & 0.381     \\
$4 P_1 $         	& 6872  & 0.416, 0.417            & 6946  & 0.429, 0.429  &  6705 & 0.376, 0.376  & 6780 & 0.384, 0.384  \\
$4 P_1' $        	& 6897  &                         & 6959  &                 & 6710 & & 6783  & \\
$4^3P_0 $       	& 6890  & 0.420                          & 6950  & 0.432   & 6693 & 0.380  & 6770 & 0.387 \\
$\theta_{4P}$   	& \multicolumn{2}{l}{31.00$^\circ$}     & \multicolumn{2}{l}{30.39$^\circ$}   & \multicolumn{2}{l}{48.6$^\circ$} & \multicolumn{2}{l}{47.4$^\circ$}  \\
\hline
$2^3G_5 $       	& 6879  & 0.407                         & 6942  & 0.419  & 6685  & 0.375  & 6766 & 0.383     \\
$2 G_4 $        	& 6875  & 0.415, 0.416              & 6938  & 0.425, 0.426   & 6672 & 0.374, 0.374  & 6756 & 0.383, 0.382 \\
$2 G_4' $       	& 6914  &                           & 6975  &                & 6764 & & 6812 & \\
$2^3G_3 $       	& 6909  & 0.424                         & 6970  & 0.432    & 6750 & 0.373  & 6801 & 0.383   \\
$\theta_{2G}$ 	& \multicolumn{2}{l}{41.76$^\circ$}     & \multicolumn{2}{l}{41.78$^\circ$}     & \multicolumn{2}{l}{47.4$^\circ$} & \multicolumn{2}{l}{47.2$^\circ$} \\
\hline
$5^3S_1 $       	& 7008  & 0.411                         & 7076  & 0.428    & 6800  & 0.372  & 6873 & 0.380  \\
$5^1S_0 $       	& 6997  & 0.416                         & 7063  & 0.426   &  6781 &  0.376  & 6858 & 0.383  \\
\hline \hline
\end{tabular}
\end{table*}

\subsection{Alternate Relativized Model}
\label{sec:arm}

The second relativized model has been developed in response to recent results from effective 
field theory and lattice gauge theory. The starting point is the general expression for the 
spin-dependent interactions in QCD as obtained from potential nonrelativistic quantum chromodynamics 
(pNRQCD) \cite{nr}. This interaction is described in terms of factorized scale-dependent Wilson 
coefficients and scale-dependent matrix elements of chromo-magnetic and -electric operators. 
The quark-antiquark interaction is modelled as the sum of a central confining term and the 
spin-dependent interaction (spin-independent corrections of order $mv^2$ also exist,
 but these are assumed to be subsumed into the central potential):
\begin{eqnarray}
V_{q\bar q}=V_{conf}+V_{SD},
\end{eqnarray}
where $V_{conf}$ is the standard Coulomb+linear scalar form
\begin{equation}
V_{conf}(r)=-C_F\frac{\alpha(r)}{r}+br.
\end{equation}

At lowest order in $\alpha_s$ the form of the spin-dependent interactions is given below \cite{BW}:
\begin{eqnarray}
V_{SD}(r) &=& \left( \frac{\bm{\sigma}_q}{4 m_q^2} +
\frac{\bm{\sigma}_{\bar q}}{4 m_{\bar q}^2} \right)\cdot {\bf L} \left( \frac{1}{r}
\frac{d V_{conf}}{d r} + \frac{2}{r} \frac{d V_1}{d r} \right)  \nonumber \\
&&+ \left( \frac{\bm{\sigma}_{\bar q} + \bm{\sigma}_q}{2 m_q m_{\bar q}} \right)\cdot {\bf L}
        \left( \frac{1}{r} \frac{d V_2}{d r} \right) \nonumber \\
&&+ {1 \over 12 m_q m_{\bar q}}\Big( 3 \bm{\sigma}_q \cdot \hat {\bf r} \,
 \bm{\sigma}_{\bar q}\cdot \hat {\bf r} -   \bm{\sigma}_q\cdot
 \bm{\sigma}_{\bar q} \Big) V_3 \nonumber \\
&&+ {1 \over 12 m_q m_{\bar q}} \bm{\sigma}_q \cdot \bm{\sigma}_{\bar q} V_4 \nonumber \\
&&\!\!\!\!\!\!\!\!\!\!\!+ {1\over 2}\left[ \left( {\bm{\sigma}_q \over m_q^2} - {\bm{\sigma}_{\bar q
}\over m_{\bar q}^2}\right)\cdot {\bf L} +
\left({\bm{\sigma}_q - \bm{\sigma}_{\bar q}\over m_q m_{\bar q}}\right)\cdot {\bf L} \right] V_5.
\label{VSD}\end{eqnarray}
Here ${\bf L} = {\bf L}_q = - {\bf L_{\bar q}}$,
$r=|{\bf r}|= |{\bf r}_q - {\bf r}_{\bar q}|$ is the quark separation
and the $V_i=V_i(m_q,m_{\bar q}; r)$ are the QCD matrix elements mentioned above.
The first four $V_i$ are order $\alpha_s$ in perturbation theory, while $V_5$ 
is order $\alpha_s^2$. 

The alternate relativized model (ARM) assumes relativistic quark kinetic energies and a running 
coupling in the Coulombic interaction. The running coupling is motivated by the persistent 
over-estimation of heavy meson decay constants \cite{LS}, and by fits to different flavor 
sectors of the meson spectrum. The form used is related to the Fourier transform of
\begin{equation}
\alpha(k) = \frac{4\pi}{\beta_0 \log\left( \exp(\frac{4\pi}{\beta_0 \alpha_0}) + \frac{k^2}{\Lambda^2}\right)},
\end{equation}
which has the expected ultraviolet behavior along with a postulated infrared fixed point. 
The leading coefficient of the QCD beta function is taken to be $\beta_0 =9$ and $\alpha_0$ 
and $\Lambda$ are free parameters to be fit to the spectrum and heavy meson decay constants.

Traditionally the forms for the potentials $V_i$ have been obtained by making nonrelativistic 
reductions of interaction kernels of the type

\begin{equation}
\frac{1}{2}\int d^3x d^3y \, \bar \psi(x) \Gamma \psi(x) \, V(x-y)\, \bar \psi(y)\Gamma \psi(y),
\end{equation}
where the Dirac matrices are typically chosen to be unity or $\gamma_0$ (these correspond to 
``scalar" or ``vector" interaction models respectively). However, QCD need not be so simple 
and we have therefore chosen to refer to lattice computations to model the interaction matrix 
elements. Direct measurement of the gluonic matrix elements reveal that $V_3$ and $V_4$ 
are short-ranged (as expected for vector interactions), while $V_1$ and $V_2$ contain 
long-ranged components \cite{KK}. However the latter do not follow the expectations 
of ``scalar confinement" \cite{ss}, rather the string tension appearing in 
$V_1$ is reduced with respect to $V_{conf}$ and there should be no long range interaction 
in $V_2$. We thus model the 
gluonic matrix elements as follows:

\begin{eqnarray}
V_1 &=& (1-\epsilon) b r \\
V_2 &=& \epsilon b r - C_F\alpha_S r \\
V_3 &=& 3 C_F\frac{\alpha_h}{r} \\
V_4 &=& C_F \alpha_h \frac{b_h^2 \textup{e}^{-b_h r}}{r} \\
V_5 &=& 0
\end{eqnarray}
Lattice results indicate \cite{KK} that $\epsilon \approx \frac{1}{4}$, which is fixed in the following.
The form of $V_4$ is based on the running coupling employed and $V_5$ has been taken to be zero 
(it was not measured in the lattice computation). The latter point is nontrivial since 
the perturbative expression for $V_5$ contains logarithms of ratios of quark masses, 
and therefore can be important in open flavor mesons. This point is discussed more in 
Ref.~\cite{ls2}. More details on the model construction are in Ref.~\cite{ess}.

A fit to 60 meson masses yields quark masses of 
$m_u =0.4585$ GeV, $m_s = 0.5919$ GeV, $m_c = 1.772$ GeV, and $m_b = 5.145$ GeV.
Potential parameters are $b= 0.1213$ GeV$^2$, $\alpha_h = 0.1536$, and $b_h = 2.138$ GeV.

\subsection{Comparison to Previous Work}

There is a long history of studying open flavor mesons in constituent quark models. Here we review some of the recent literature, noting similarities and differences with the current approach.  

A notable early contribution was due to Di Pierro and Eichten, who examined the $D$, $D_s$, $B$, and $B_s$ spectra with a constituent quark model based on the Dirac equation for the light 
quark
in the potential generated by the heavy quark\cite{DiPierro:2001dwf}.  Strong decays of these states by pseudoscalar meson emission were then  computed with the chiral quark model of Manohar and Georgi\cite{gm}.

The unusual mass of the $D_s(2317)$ generated a series of papers on open flavor meson spectroscopy. Here we mention that of Close and Swanson, who computed the $D$ and $D_s$ spectra using a simple  nonrelativistic quark model with a central Cornell potential and a smeared hyperfine interaction\cite{Close:2005se}. Strong decays were computed with the $^3P_0$ model with SHO mesonic wavefunctions and effective meson widths (as is done here). Radiative transitions, including those from molecular states, were also presented.

The $D_s(2317)$ also motivated the analysis of Ref. \cite{ls2}. The novel feature of this investigation was the use of $O(\alpha_s^2)$ spin-dependent corrections to the static interquark potential -- in particular flavor-dependence is introduced via logarithmic dependence on quark mass. This can have strong effects on the open-flavor spectrum and can shift the nominal scalar meson mass down by roughly 100 MeV.  This paper also examined relativistic effects (due to light quark spinors) on radiative transitions, and found large shifts.

Alternative relativistic (or  ``relativized") models also appeared in the late 2000s. For example, Matsuki \textit{et al.}, who used scalar and vector interaction kernels in a Dirac equation with a Foldy-Wouthuysen reduction of the heavy quark portion of the interaction\cite{Matsuki:2007zza}. Related work by Ebert \textit{et al.} appeared at about this time as well\cite{Ebert:2009ua}. These authors used relativistic kinetic energies and vector, chromomagnetic, and scalar Dirac structures in the 
interaction
kernels to obtain the $D$, $D_s$, $B$, and $B_s$ spectra. Finally, Devlani and Rai made an $O(p^6)$ reduction of the quark kinetic energy and supplemented this 
with
a Cornell central potential and the Breit spin-dependent interaction\cite{Devlani:2012zz}. 
Unfortunately, the authors appeared to have employed an unregulated delta function in the latter interaction, which is not acceptable in quantum mechanical problems (evidently this problem was not discovered because the spectrum was obtained with a simple variational calculation).
The first two of these papers did not examine radiative or strong transitions; radiative transitions were computed in the latter.

More recently, the chiral quark model has been revisited for the strong decays of open flavor mesons\cite{Xiao:2014ura}. This work employed SHO mesonic wavefunctions with a fixed Gaussian width. Two additional papers have computed $B$ and  $D$ meson spectra with the Godfrey-Isgur model\cite{Sun:2014wea,Ferretti:2015rsa}. They have also calculated strong decays using a very similar method to that presented here; the main difference is that the latter group incorporated a quark form factor in the $^3P_0$ vertex to suppress high energy decay modes and used SHO mesonic wavefunctions with a single Gaussian width. A similar computation of the $B$ and $D$ spectra was made with a model with relativistic quark kinetic energy and an interaction with a running coupling and smeared delta functions\cite{Liu:2015lka}. Decays were not considered. Finally, Liu \textit{et al.} have made a Foldy-Wouthuysen reduction of the instantaneous approximation to the Bethe-Salpeter equation and obtained the $D$, $D_s$, $B$, and $B_s$ spectra\cite{Liu:2016efm}.

Shortly after this work appeared, a computation that used the model of Ref. \cite{ls2} 
was submitted to the arXiv \cite{Lu:2016bbk}. The authors computed $B$ and $B_s$ spectra 
and strong decays with the $^3P_0$ model using full mesonic wavefunctions. A factor of 
$m_u/m_s$ was applied to the $^3P_0$ coupling to suppress strange quark production. 
This factor can be justified by considering the quark pair production interaction to 
be proportional to $\int \bar \psi \psi${\cite{Ackleh:1996yt}, however this is a model 
assumption, and we must resort to experiment to decide the issue.

\section{Radiative Transitions}

\subsection{E1 Transitions}

Radiative transitions could play an important role in the discovery 
and identification of $B$ and $B_s$ states.    They are sensitive to the internal
structure of states, in particular to $^3L_L - {}^1L_L$ mixing for states with $J=L$.  
In this section we calculate the electric dipole (E1) and 
magnetic dipole (M1) radiative widths.
The partial width for an E1 radiative transition between
states in the nonrelativistic quark model is given by 
\cite{Kwo88}
\begin{widetext}
\begin{equation}
\Gamma( 
{n}\, {}^{2{S}+1}{L}_{J} 
\to 
{n}'\, {}^{2{S}'+1}{L}'_{{J}'}  
+ \gamma) 
 =  \frac{4}{3}\,  \langle e_Q \rangle^2 \, \alpha \,
k^3 \,   
C_{fi}\,
\delta_{{S}{S}'} \delta_{{L}{L'\pm1}} \, 
|\,\langle 
{n}'\, {}^{2{S}'+1}{L}'_{{J}'} 
|
\; r \; 
|\, 
{n}\, {}^{2{S}+1}{L}_{J}  
\rangle\, |^2  
\, 
\end{equation}
\end{widetext}
where 
\begin{equation}
\langle e_Q \rangle = {{m_b e_q - m_q e_{\bar{b}} }\over{m_q +m_b}}
\end{equation}
where $q = u,d,s$, we use the quark masses $m_u=m_d=0.220$~GeV, $m_s=0.419$~GeV and $m_b=4.977$~GeV, $e_u= +2/3$ and $e_d=e_s=e_b= -1/3$
are the quark charges in units of $|e|$,
$\alpha$ is the fine-structure constant,
$k$ is the photon's energy, and $C_{fi}$ is given by
\begin{equation}
C_{fi}=\hbox{max}({L},\; {L}') (2{J}' + 1)
\left\{ { {{L}' \atop {J}} {{J}' \atop {L}} {{S} \atop 1}  } \right\}^2 .
\end{equation}
where $\{ {\cdots \atop \cdots} \}$ is a 6-$j$ symbol.
The matrix elements $\langle {n'}^{2{S}'+1}{L}'_{{J}'} |\; r \; 
| n^{2{S}+1}{L}_{J}  \rangle$ 
are given in Tables \ref{tab:B_12S}-\ref{tab:Bs_1G} where applicable and were evaluated 
using the wavefunctions given by the relativized quark model \cite{Godfrey:1985xj}.
Relativistic corrections are implicitly included in these E1 
transitions through Siegert's theorem \cite{Sie37,McC83,Mox83}
by including spin-dependent interactions in the Hamiltonian used to 
calculate the meson masses and wavefunctions.   
The E1 radiative widths are given in Tables \ref{tab:B_12S}-\ref{tab:Bs_1G} where applicable.

\subsection{M1 Transitions}

Radiative transitions which flip spin are described by magnetic dipole
(M1) transitions.  The rates for magnetic dipole transitions between $S$-wave heavy-light 
bound states are given in the nonrelativistic approximation 
by \cite{JDJ,Nov78}
\begin{widetext}
\begin{equation}
\Gamma(n^{2S+1}L_J  \to   {n'} ^{2S'+1}L_{J'} + \gamma)  
  =   {\alpha \over 3} k^3 (2J'+1) \delta_{S,S'\pm 1}
\left| {e_q \over m_q}\langle f | j_0 \left({m_b \over m_q+m_b} kr \right) | i \rangle + {e_b \over m_b}\langle f | j_0 \left({m_q \over m_q+m_b} kr \right) | i \rangle \right|^2
\end{equation}
\end{widetext}
where $e_q$, the quark charges, and $m_q$, the quark masses, were given above, $L=0$
for $S$ waves and $j_0(x)$ is the spherical Bessel function.

The M1 widths and overlap integrals are given in Tables 
\ref{tab:B_12S}-\ref{tab:Bs_1G} where applicable. 
Transitions in which the principle quantum number 
changes are refered to as hindered transitions, which
are not allowed in the non-relativistic limit due to the orthogonality 
of the wavefunctions. 
M1 transitions, especially hindered transitions,
are notorious for their sensitivity to relativistic 
corrections \cite{m1}.   
In our calculations the wavefunction orthogonality is broken 
by including a smeared  hyperfine interaction directly in the 
Hamiltonian so that the $^3S_1$ and $^1S_0$ states have slightly 
different wavefunctions. 
Ebert {\it et al.}  are more rigorous in how they include relativistic 
corrections \cite{ebert03} but 
to improve the $J/\psi \to \eta_c \gamma$ result they modify the 
confining potential  by making it a 
linear combination of Lorentz vector and Lorentz scalar pieces.

The E1 and M1 radiative widths are given in Tables \ref{tab:B_12S}-\ref{tab:Bs_1G}
when they are large enough that they might be observed.  The predicted masses given in
Tables \ref{tab:masses}-\ref{tab:masses2} 
are used for all states under the assumption that predicted masses are expected
to be shifted by
comparable amounts from their measured masses leaving the phase 
space to remain approximately correct. 

Measuring radiative widths can help identify newly observed states and in addition, 
given the sensitivity of radiative transitions to details of the models,
precise measurements of electromagnetic transition rates would 
provide stringent tests of the various calculations.

\section{Strong Decays}

We use the $^3P_0$ model  \cite{Micu:1968mk,Le Yaouanc:1972ae,Ackleh:1996yt,Blundell:1995ev,Barnes:2005pb} 
to calculate all kinematically allowed strong decay widths for the $B$ and $B_s$ meson states 
listed in 
Tables \ref{tab:masses}-\ref{tab:masses2}.
The masses shown are the theoretical values calculated 
using the Godfrey-Isgur relativized quark model \cite{Godfrey:1985xj}.  We use harmonic 
oscillator wave functions with the effective oscillator parameter, $\beta_{eff}$, obtained by 
equating the rms radius of the harmonic oscillator wavefunction for the specified $(n,l)$ 
quantum numbers to the rms radius of the wavefunctions calculated using the relativized 
quark model of Ref.~\cite{Godfrey:1985xj}.  
The effective harmonic oscillator wave function parameters, $\beta_{eff}$, 
that
we use in our 
calculations are listed in 
Tables \ref{tab:masses}-\ref{tab:masses2}.
For the light mesons, we use a common value 
of $\beta_{eff} = 0.4$~GeV
(see below) and the experimental masses as input, given in Table \ref{tab:udsparams}.  
For the constituent quark masses in our calculations of both the meson masses and of the 
strong decay widths, we use $m_b=4.977$~GeV, $m_s=0.419$~GeV and $m_q=0.220$~GeV.  
Finally, we use ``relativistic phase space'' as described in Ref.~\cite{Geiger:1994kr,Blundell:1995ev}.
We use the calculated bottom and bottom-strange meson masses listed in 
Tables \ref{tab:masses}-\ref{tab:masses2}. 
For the light mesons we used the measured masses listed in Table \ref{tab:lightmasses}.  
Details regarding the notation and 
conventions used in the $^3P_0$ model calculations are given in the appendix of \cite{Godfrey:2015}.

\begin{table}
\caption{Light meson masses and effective harmonic oscillator parameters, $\beta_{eff}$, 
used in the calculation of strong decay widths.  The experimental values of the masses 
are taken from the Particle Data Group (PDG) \cite{Olive:1900zz}.  The input value of 
the $\pi$ mass is the weighted average of the experimental values of the $\pi^0$ and $\pi^\pm$ 
masses, and similarly for the input values of the $K$ and $K^*$ masses.  
All $\beta_{eff}$ values are taken to be 0.4~GeV for the light mesons.
\label{tab:lightmasses}}
\begin{ruledtabular}
\begin{tabular}{lllll}
Meson			&	State			&  $M_{input}$ 		& $M_{exp}$  \cite{Olive:1900zz}	& $\beta_{eff}$ 	\\
				&				& (MeV)			& (MeV)							&  (GeV)			\\
\hline \hline
$\pi$				&	$1^1S_0$		&	138.8877		&	$134.8766\pm0.0006~(\pi^0)$, 		&	0.4			\\
				&				&				&   $139.57018\pm0.00035~(\pi^\pm)$	&				\\
$\eta$			&	$1^1S_0$		&	547.862		&	$547.862\pm0.018$ 				&	0.4			\\
$\eta^{\prime}$		&	$1^1S_0$		&	957.78		&	$957.78\pm0.06$ 				&	0.4			\\
$\rho$			&	$1^3S_1$		&	775.26		&	$775.26\pm0.25$ 				&	0.4			\\
$\omega$			&	$1^3S_1$		&	782.65		&	$782.65\pm0.12$ 				&	0.4			\\
$\phi$			&	$1^3S_1$		&	1019.461		&	$1019.461\pm0.019$			&	0.4			\\
\hline
$K$				&	$1^1S_0$		&	494.888		&	$497.614\pm0.024~(K^0)$,   		&	0.4			\\
				&				&				&   $493.677\pm0.016~(K^\pm)$		&				\\
$K^*$			&	$1^3S_1$		&	894.36		&	$895.81\pm0.19~(K^{*0})$, 		&	0.4			\\
				&				&				&   $891.66\pm0.26~(K^{*\pm})$ 		&				\\
\end{tabular}
\end{ruledtabular}
\label{tab:udsparams}
\end{table}

Typical values of the parameters $\beta_{eff}$ and $\gamma$ are found from fits to light meson 
decays \cite{Close:2005se,Blundell:1995ev,Blundell:1996as}. The predicted widths are fairly
insensitive to the precise values used for $\beta_{eff}$ provided $\gamma$ is appropriately rescaled.  
However $\gamma$ can vary as much as 30\% and still give reasonable overall fits of
 light meson decay widths \cite{Close:2005se,Blundell:1996as}. This can result in factor 
 of two changes to predicted widths, both smaller or larger.  In our calculations of 
 $D$ and $D_s$ meson strong decay widths from \cite{Godfrey:2014fga} and \cite{Godfrey:2015dva}, we used a value of $\gamma=0.4$, 
 which has also been found to give a good description of strong decays of charmonium 
 \cite{Barnes:2005pb,Close:2005se}.  We adopt the same value of $\gamma = 0.4$ in our 
 calculations of $B$ and $B_s$ meson strong decays.  The resulting partial widths are 
listed in Tables~\ref{tab:B_12S}-\ref{tab:Bs_1G}.  We include more complete sets of decays
in a supplementary file that includes decays not included in the paper because we felt 
that their BR's were too small to likely be observed. To make our results
as comprehensive as possible the supplementary file also
includes tables of strong decays for the $5S$, $4P$, $3D$, $2F$ and $2G$ states.

\begin{table*}


\section{Model Sensitivity}

The results of Tables \ref{tab:B_12S}--\ref{tab:Bs_1G} are presented in terms of Godfrey-Isgur model masses and wavefunctions. Rather than double the number of tables by giving the analogous ARM results, we have chosen to represent model sensitivity graphically. For example, Fig. \ref{fig:mass} is a scatter plot of Godfrey-Isgur versus ARM masses. A close correspondence is evident, although ARM masses tend to be lower than GI masses higher in the spectrum due to the different string tensions employed in the models.

\begin{figure}[h]
\includegraphics[width=9.5cm,angle=0]{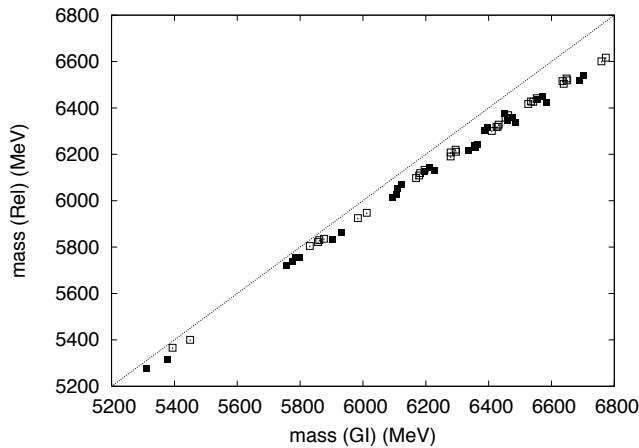}
\caption{Mass predictions for Godfrey-Isgur (GI) and ARM (Rel) models. Solid 
symbols are $B$ states; open symbols are $B_s$ states.}
\label{fig:mass}
\end{figure}

The similarity of the mass predictions is 
also reflected in the high correlation of the predicted  
strong decay widths of the two models, shown in Fig. \ref{fig:dec}. 
An alternate measure of the model sensitivity is to calculate the relative difference
of the predicted strong decay widths which we take to be the difference between the GI
predictions and the `alternate model' divided by the GI prediction.  For the predictions 
of the alternate model, data were generated under a variety of conditions; specifically (i) full 
ARM predictions, (ii) ARM wavefunctions, (iii) ARM wavefunctions and quark masses, 
(iv) ARM wavefunctions and meson masses.  For the purposes of illustration a representative
set of decays were used to construct the frequency histogram of the relative difference 
of predicted strong decay widths shown in Fig. \ref{fig:reldev}.
One observes that the distribution is peaked 
at zero deviation, while the average deviation is 14\%.

\begin{figure}[h]
\includegraphics[width=9.5cm,angle=0]{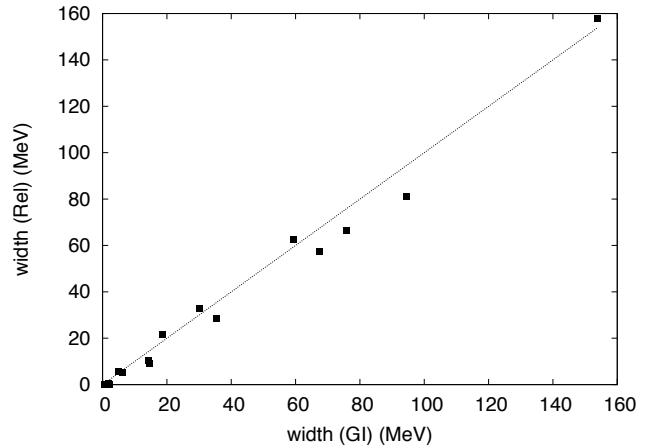}
\caption{Selected Strong Decay Width Predictions for Godfrey-Isgur (GI)  and AR (Rel) models.}
\label{fig:dec}
\end{figure}

\begin{figure}[h]
\includegraphics[width=9.5cm,angle=0]{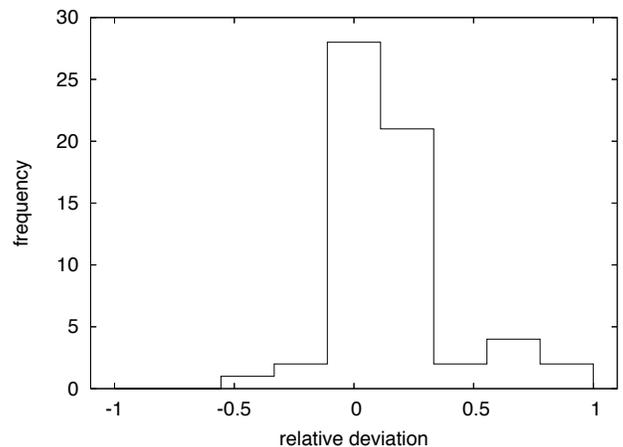}
\caption{Relative deviation of strong decay predictions.}
\label{fig:reldev}
\end{figure}

Radiative transitions can be very sensitive to model details. For example, the size 
of hindered M1 transitions depends crucially on assumed wavefunctions. Similarly, 
radiative transition rates for heavy-light mesons involve differences of light and 
heavy quark amplitudes, and these differences depend strongly 
on whether the light quark amplitude is considered in the nonrelativistic limit or 
not \cite{ls2}. It is therefore perhaps not surprising that the GI and AR models 
deviate somewhat in their predicted E1 transition rates, as shown in Fig.~\ref{fig:E1}. 
Fig. \ref{fig:relE1} shows the related frequency histogram, calculated as described above,
and indicates that the GI E1 
predictions tend to be approximately 50\% larger than those of the AR model. 

We find these results reasonably reassuring. It appears that constituent quark models that 
have been tuned to a wide range of hadrons provide similar predictions for hadronic properties. 
Perhaps the most interesting deviation seen here is the systematically lighter predictions of 
excited $B$ and $B_s$ states by the AR model with respect to the GI model. This is almost 
certainly due to the differing string tensions employed in the two models. Certainly, 
the `stiff' string tension of the GI model is preferred by lattice 
Wilson loop computations
and bottomonium spectroscopy. 
However, the smaller string tension of the ARM fits lighter 
mesons better. It will be interesting to see which is preferred in the description of heavy-light mesons.

\begin{figure}[h]
\includegraphics[width=9.5cm,angle=0]{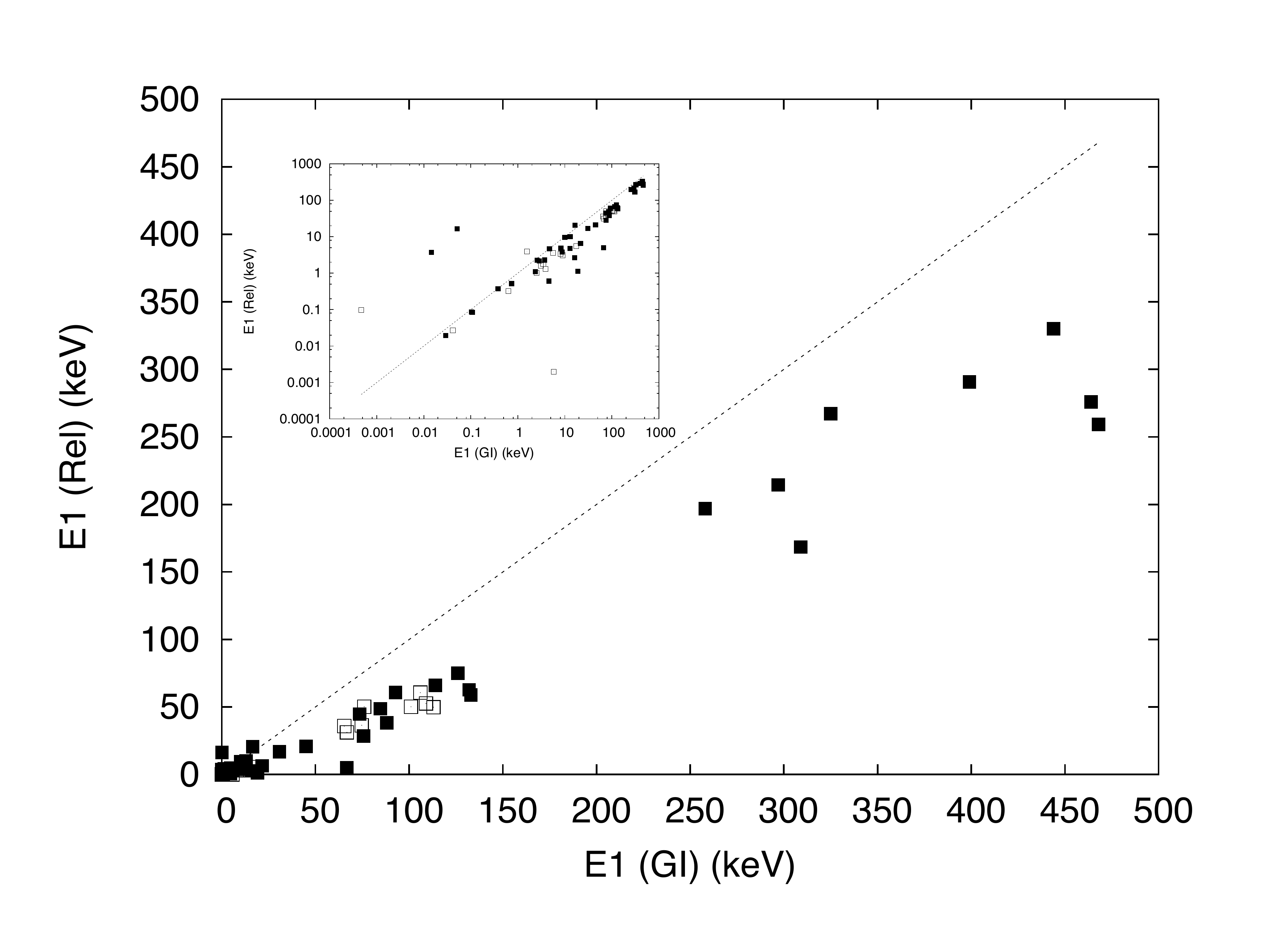}
\caption{E1 transition rate predictions for Godfrey-Isgur (GI) and AR (Rel) models. Solid symbols are $B$ states; open symbols are $B_s$ states. Inset: log-log scale.}
\label{fig:E1}
\end{figure}

\begin{figure}[h]
\includegraphics[width=9.5cm,angle=0]{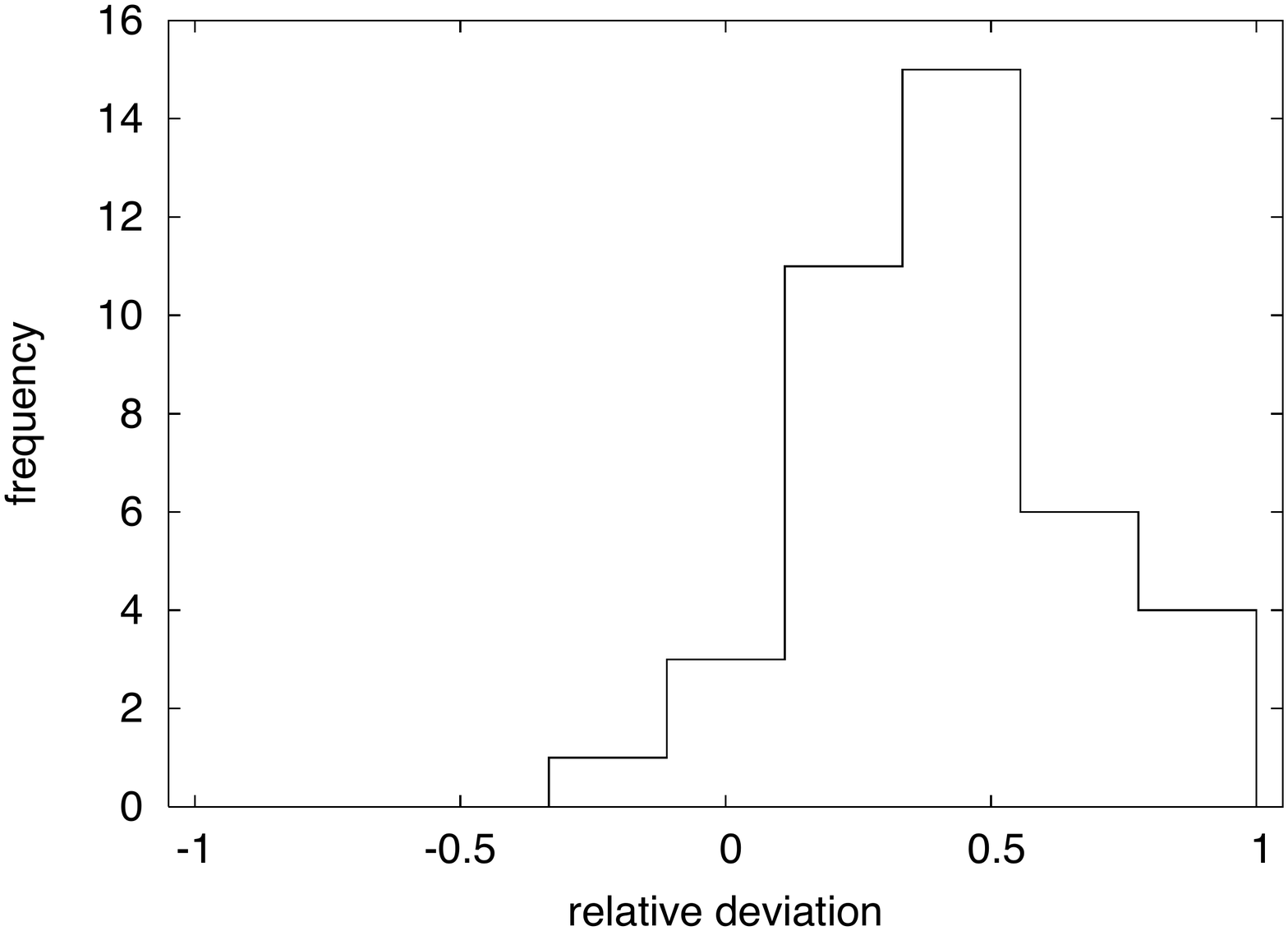}
\caption{Relative deviations of E1 radiative transition rates.}
\label{fig:relE1}
\end{figure}

\section{Classification of the Observed Bottom Mesons}

The experimental knowledge of excited $B$ mesons is rather limited.  However recent 
measurements by the LHCb collaboration \cite{Aaij:2015qla} have demonstrated the
potential to extend our knowledge of these states considerably.
In the following section we will discuss the excited $B$ mesons we believe are most
likely to be observed in the near future and how they might be observed.  In this 
section we will summarize the existing situation, comment on spectroscopic assignments
and suggest measurements that would improve our understanding of these states.  A summary of the experimental status of the excited $B$ mesons is presented in Table~\ref{tab:obs_bottom}.

Starting with the $B$ mesons,  four members of the $1P$ multiplet are expected to be
seen around 5750~MeV, comprised of a doublet of two relatively narrow states and a doublet
of two relatively broad states.  The $B_1(5727)$ and $B_2^*(5739)$ with widths of
30~MeV and 24~MeV respectively (we quote recent results from LHCb \cite{Aaij:2015qla}
which have smaller errors than listed in the PDG \cite{Olive:1900zz}) have measured 
properties that are in qualitative agreement with our predictions.  The
GI model overestimates the $B_1$ and $B_2^*$ masses by approximately $50$~MeV.  However the GI 
model also overestimates the 
$B$ and $B^*$ masses 
by a similar amount so it is reasonable to assume
that the GI model in general overestimates $B$ masses by $\sim$50~MeV.  In contrast, the
ARM mass predictions agree with experiment within approximately $10$~MeV.  However, while the 
model predicts a doublet of two relatively narrow states, the predicted widths are quite a bit
smaller than the measured widths; 7~MeV vs 30~MeV for the $B_1$ and 12~MeV vs 24~MeV
for the $B_2^*$.  In general we would not be surprised if our width predictions were off
by a factor of 2 so would consider the $B_2^*$ width prediction to be acceptable also taking
into account experimental error.  However the discrepancy of the $B_1$ width prediction
is most likely due to the sensitivity of the prediction to the $^3P_1-^1P_1$ mixing angle 
($\theta_{1P}$).  
Even a small change in mixing angle would 
have a dramatic effect on the width prediction
and in fact we could use this information to constrain $\theta_{1P}$.

The $B(5970)$ has been seen in the $B\pi$ final state by the CDF collaboration with 
a mass of $5961\pm 13$~MeV for the $B(5970)^+$ and $5977\pm 13$~MeV for the $B(5970)^0$
with widths of $60\pm 50$~MeV and $70\pm 42$~MeV respectively \cite{Olive:1900zz}.  
The only $B$ mesons that
are close in mass to the $B(5970)$ are the $B^*(2^3S_1)$ and $B(2^1S_0)$ states. However, the
$B(2^1S_0)$ cannot decay to a $B\pi$ final state, leaving only the  $B^*(2^3S_1)$
as a possible candidate.  The 
GI model predicts the mass of the $B^*(2^3S_1)$  to be 5933~MeV 
although  as previously noted we believe
that the GI model overestimates $B$ masses by approximately $50$~MeV.  The AR model predicts the mass
of this state to be  5864~MeV.  
Thus the observed mass appears to be too large to 
be identified with the $B^*(2^3S_1)$  state.  
However the predicted width for the  $B^*(2^3S_1)$ 
is 108~MeV   which, given both
the large experimental errors and theoretical uncertainties, is consistent with the
measured widths.  So while the $B(5970)$ might be identified with the $B^*(2^3S_1)$, 
given the large discrepancy between the predicted and observed mass we wait for further
measurements of this state before making a conclusion.

Recently LHCb \cite{Aaij:2015qla}
reported the observation of new states, the $B_J(5840)$ and the $B_J(5960)$
with masses and widths of $M(B_J(5840))=5857$~MeV, $\Gamma (B_J(5840))=175.9$~MeV
and $M(B_J(5960))=5967$~MeV, $\Gamma (B_J(5960))=73$~MeV where we have averaged the masses
and widths for the two observed charge states and do not quote experimental errors as
they are quite large and the extracted values are very model dependent.  LHCb 
suggests these may be the $B(2^1S_0)$ and $B^*(2^3S_1)$ states.  This can be compared to
the predicted masses (from ARM) and widths for the two states of 
$M=5834$~MeV and $\Gamma = 95$~MeV for the $B(2^1S_0)$ and $M=5864$~MeV and $\Gamma = 108$~MeV
for the $B^*(2^3S_1)$. The predicted properties of the $B(2^1S_0)$ are consistent with those of 
$ B_J(5840)$ within experimental and theoretical uncertainties and while the predicted
$B^*(2^3S_1)$ width is consistent with that of the $B_J(5960)$, the predicted mass is about 100~MeV
too low.  All things considered, these new states can be identified with the $2S$ $B$ mesons
but ideally more measurements are needed to support this conclusion.  A final comment
is that the properties of the $B(5970)$ seen by CDF are consistent with the properties of
the $B_J(5960)$ seen by LHCb.

The two remaining excited bottom mesons that have been observed are the 
bottom-strange states: the $B_{s1}(5830)^0$ 
with $M=5828.78\pm 0.35$~MeV and $\Gamma = 0.5 \pm 0.4$~MeV
and the $B_{s2}^*(5840)^0$ with $M= 5839.83 \pm 0.19$~MeV and $\Gamma = 1.47 \pm 0.33$~MeV \cite{Olive:1900zz}.  
Further, for the $B_{s2}^*$, $\Gamma (B^{*+} K^-)/\Gamma(B^+ K^-) = 0.093 \pm 0.018$.  
We compare
these to the properties of the $B_{s1}(1P)$ with $M=5822$~MeV and $\Gamma = 0.11$~MeV
and the $B_{s2}^*(1^3P_2)$ with $M=5836$~MeV and $\Gamma = 0.78$~MeV where we quote the mass
predictions of the AR model.  The predicted masses are in good agreement with the measured masses
and the predicted width of the 
$B_{s2}^*(1^3P_2)$ is in acceptable agreement with the
$B_{s2}^*(5840)^0$ measured width.  However, as in the case of the $B_1(5727)$ we again 
surmise that the discrepancy is due to the sensitivity of the $B_{s1}(1P)$ width to the 
$^3P_1 -^1P_1$ mixing angle.  
The $B_{s1}(1P)$ is also close to the $B^*K$ threshold so is very sensitive to phase space.  
As shown in Table \ref{tab:Bs_1P}, there are no kinematically allowed strong decay modes for the $B_s(1P_1)$ and $B_s(1P_1^\prime)$ when using their predicted Godfrey-Isgur masses as input.  However, using the measured mass for the $B_{s1}(1P)$ state as input to the $^3P_0$ model calculations opens up the $B_s(1P_1) \to B^*K$ decay mode with a partial width of 0.331 MeV.  

To summarize, the information on excited bottom mesons is limited.  For both the bottom
and bottom-strange mesons, two narrow states have been seen and their properties are 
consistent with those of the $1^3P_2$ and $1P_1$ quark model states.  Two more states
have been observed, most recently by the LHCb collaboration \cite{Aaij:2015qla}, 
but the experimental uncertainties
are quite large so although these states may be the $B(2^1S_0)$ and $B^*(2^3S_1)$ states
more data is needed to confirm these identifications.  However, the recent LHCb results 
show the promise that LHCb holds in opening up new frontiers in bottom meson spectroscopy.
In the next section we examine the $B$ meson landscape and suggest promising avenues for
finding new excited bottom mesons.

\begin{table*}[t]
\caption{Summary of excited bottom mesons.  Unless otherwise stated we quote values from
the Particle Data Group \cite{Olive:1900zz}.
\label{tab:obs_bottom}}
\begin{tabular}{llllll} \hline \hline
State 			&  $J^P$	& Observed Decays & Mass (MeV) 	& Width (MeV) 	  & References  \\ \hline
$B_1(5721)^+$	& 	& $B^{*0}\pi^+$	& $5726.8^{+3.2}_{-4.0}$	& $49^{+12 +2}_{-10-1.3}$ &  \\
				& 	& 				& $5725.1\pm 1.8 \pm 3.1 \pm 0.17 \pm 0.4$	& $29.1 \pm 3.6 \pm 4.3$ & LHCb\cite{Aaij:2015qla}  \\
$B_1(5721)^0$	& 	& $B^{*+}\pi^-$	& $5724.9\pm 2.4$	& $23 \pm 3\pm 4$ &  \\
				& 	& 				& $5727.7\pm 0.7 \pm 1.4 \pm 0.17 \pm 0.4$	& $30.1 \pm 1.5 \pm 3.5$ & LHCb\cite{Aaij:2015qla}  \\
$B_2^*(5747)^+$	& 	& $B^{0}\pi^+$	& $5736.9^{+1.3}_{-1.6}$	& $11^{+4+3}_{-3-4}$ &  \\
				& 	& 				& $5737.20\pm 0.72 \pm 0.40 \pm 0.17 $	& $23.6 \pm 2.0 \pm 2.1$ & LHCb\cite{Aaij:2015qla}  \\
				&	& 			&		& $\Gamma (\to B^{*0}\pi^+)/\Gamma(\to B^0 \pi^+) = 1.0 \pm 0.5 \pm 0.8$ &  LHCb\cite{Aaij:2015qla}\\
$B_2^*(5747)^0$	& 	& $B^{+}\pi^-$, $B^{*+}\pi^-$	& $5739^\pm 5$	& $22^{+3+4}_{-2-5}$ &  \\
				&	& 			&		& $\Gamma (\to B^{*+}\pi^-)/\Gamma(\to B^+ \pi^-) = 1.10 \pm 0.42 \pm 0.31$ &  \\
				&	&		& $5739.44\pm 0.37 \pm 0.33 \pm 0.17$ & $24.5\pm 1.0 \pm 1.5$ & LHCb\cite{Aaij:2015qla}  \\
				& 	& 				& $5739.44\pm 0.37 \pm 0.33 \pm 0.17 $	& $24.5 \pm 1.0 \pm 1.5$ & LHCb\cite{Aaij:2015qla}  \\
				&	& 			&		& $\Gamma (\to B^{*+}\pi^-)/\Gamma(\to B^+ \pi^-) = 0.71 \pm 0.14 \pm 0.30$ &  LHCb\cite{Aaij:2015qla}\\
$B(5970)^+$	& 	& $B^{0}\pi^+$	& $5961\pm 13$	& $60^{+30}_{-20}\pm 40$ &  \\
$B(5970)^0$	& 	& $B^{+}\pi^-$	& $5977\pm 13$	& $70^{+30}_{-20}\pm 30$ &  \\
\hline
$B_{s1}(5830)^0$	& 	& $B^{*+}K^-$	& $5828.78 \pm 0.35 $	& $0.5 \pm 0.3 \pm 0.3 $ &  \\
$B_{s2}^*(5830)^0$	& 	& $B^{+}K^-$	& $5839.83 \pm 0.19 $	& $1.47 \pm 0.33  $ &  \\
				&	& 			&		& $\Gamma (\to B^{*+}K^-)/\Gamma(\to B^+ K^-) = 0.093 \pm 0.013 \pm 0.012$ &  \\
%
%
\hline
\hline
\end{tabular}
\footnotetext[1]{We quote the results from the isobar analysis.}
\end{table*}

\section{Experimental Signatures and Search Strategies}

We expect that the excited $B$ mesons most likely to be observed in the near future
are those that are relatively narrow and that have a large branching ratio (BR) to a simple final state such as
$B\pi$, $B^*\pi$, $BK$, or $B^* K$.  There has been great recent progress in finding
excited charm mesons in the analogous $D\pi$ final states \cite{Godfrey:2015dva}. In this section
we examine Tables \ref{tab:B_12S}-\ref{tab:Bs_1G} to identify states that meet these criteria.

Starting with the $B$ and $B_s$ $1P$ multiplets the $1P_1$ and $1^3P_2$ states have already been seen.  
Their predicted  properties are in reasonable agreement with the measured properties
 within the theoretical uncertainties.  The hallmark of these states is that they 
 have relatively small total widths with large BR's to simple final states.  We expect 
 the total widths for the $j_q=1/2$ doublet to be considerably larger but still not 
 so large that an observable signal in a simple final state should be seen.  
Specifically, we calculate the total width for the $B(1^3P_0)$ to be 154~MeV decaying 
almost 100\% of the time to $B\pi$ (with a small BR to $B^*\gamma$), 
and $\Gamma(B(1P_1'))=163$~MeV decaying almost 100\% of the time to $B^*\pi$.
Similarly we calculate $\Gamma (B_s(1^3P_0))=138$~MeV decaying predominantly to $BK$, and
the $B_s(1P_1')$ is below $B^*K$ threshold so is expected to be quite narrow with its
dominant BRs to $B_s \gamma$ and $B_s^*\gamma$.  The ratio of these two BR's would
determine the $^3P_1 - ^1P_1$ mixing angle.  For these predicted total widths it should be
possible to observe the missing $B$ and $B_s$ $1P$ states in the near future.  The challenge
will be disentangling overlapping states but this should be possible with sufficient statistics.

Next highest in mass are the $2S$ states.  The $B(2^3S_1)$ is predicted to have 
a mass of 5864~MeV, and $\Gamma = 108$~MeV with BR$(B\pi)=33\%$ and BR$(B^*\pi)=63\%$.
The properties of the $B(2^1S_0)$ are $M=5834$~MeV, $\Gamma=95$~MeV with BR$(B^*\pi)\simeq 100\%$.
In both cases we quote the ARM mass prediction.
There is evidence that LHCb has seen these states but as discussed above more data is 
needed to confirm this.  For the analogous bottom-strange  states we find
for the $B_s(2^3S_1)$ $M=5948$~MeV, $\Gamma = 114$~MeV with BR$(BK)=38.3\%$ and
BR$(B^*K)=58.4\%$, and for the $B_s(2^1S_0)$ $M=5925$~MeV, $\Gamma = 76$~MeV with 
BR$(B^*K) \simeq 100\%$.  These states are relatively narrow with large BR's to simple final
states so it should be possible to observe them in the near future.

As we move to higher mass states the situation becomes more complicated.  In general 
the mass multiplets are closer together with relatively small splittings within multiplets
$({\cal O}(10 \hbox{ MeV}))$,  and many of the states become broader due to the 
increased phase space resulting in many overlapping resonances.  To disentangle this will require
higher statistics to measure the spins.  LHCb has demonstrated their ability to accomplish this
with spin measurements in the charm meson sector.  However,  not all states are broad 
due to angular momentum suppression in decays so it should still be possible to find
some of these excited states.  In what follows we will focus on the states most likely to
be found first. 

The $1D$ multiplets are next highest in mass.  They consist of a narrow doublet and a broad doublet.
The narrow $B$ doublet consists of the $1D_2$ with total width 23~MeV
with BR$(B^* \pi)=87\%$ and the $1^3D_3$ with 
total width 31~MeV with 
BR$(B\pi)=46\%$ and BR$(B^*\pi)=46\%$.  These
two states should have strong signals in their dominant decay modes.  However, because
these two states are close in mass it will likely require an angular distribution analysis to 
distinguish the $1D_2$ from the 
$1^3D_3$ in the $B^*\pi$ final state. Nevertheless it should 
be possible to observe the two narrow $1D$ states.  In contrast, for the two broad states, 
$1^3D_1$ and $1D_2'$,   despite the fact that the $1D_2'$ 
is expected to have a large BR of 45\% to $B^*\pi$ and the $1^3D_1$ of 30\% to $B\pi$, because 
they are expected to have total widths of approximately $200$~MeV it will likely be difficult 
to extract a strong signal.  These observations equally apply to the
$B_s (1D)$ states.   The narrow $B_s(1D_2)$ and $B_s(1^3D_3)$ states 
have widths of 16~MeV and 26~MeV respectively.  The $B_s(1D_2)$ has a  BR to $B^*K$ of
97\% while the  $B_s(1^3D_3)$ has BR's to $BK$ of 53\% and to $B^*K$ of 43\%.  In contrast, the 
broad $D_s(1^3D_1)$ has a width of 183~MeV with a BR to $BK$ of 59\% and to $B^*K$ of 29\%
while the broad $B_s(1D_2')$ has a width of 194~MeV with BR to $B^*K$ of 88\%.  We conclude
that the narrow states will produce strong signals although sitting in broad backgrounds 
so that measuring their spins and parities will be helpful in identifying them.

Next in mass are the $2P$ multiplets.  All members of the nonstrange bottom $2P$ 
multiplet are relatively broad with $\Gamma \sim 200$~MeV which is significantly greater than the
intra-multiplet mass splitting.  The largest BR's are mainly to more complicated final states,
for example BR$(B(2^3P_0)\to B(1P_1)\pi)=50\% $ and BR$(B(2P_1')\to B(1^3P_2)\pi)=41\% $.  The
most promising possibility is to study the $B\pi$ final state which only the $B(2^3P_0)$
and $B(2^3P_2)$ can decay to with BR's of 10.5\% and 8.4\% respectively.  Because the
expected widths are so much larger than the splitting, distinguishing the states will 
require an angular momentum analysis to identify the two states.  The $B(2P_1)$ and 
$B(2P_1')$ both decay to $B^*\pi$ with BR's of 30\% and 9\% respectively.  But again,
because they are overlapping resonances more information would be needed to distinguish 
the two states, such as BR's to other final states such as $B\rho$ and $B^*\rho$, although 
this would be difficult because the $B(2^3P_0)$ and $B(2^3P_2)$ 
also decay to 
$B^*\rho$ with BR's of approximately $20\%$ and $40\%$ respectively.  We conclude that only the 
$B(2^3P_0)$  and $B(2^3P_2)$  might be observed in the foreseeable future.  

In contrast, the $2P$ $B_s$ mesons divide into two relatively narrow states,
$B_s(2^3P_0)$ and $B_s(2P_1')$, and two relatively
broad states, $B_s(2P_1)$ and $B_s(2^3P_2)$.  The $B_s(2^3P_0)$ has a total width of 71~MeV and 
BR to $BK$ of 44\% and the $B_s(2P_1')$ has a total width of 79~MeV with a BR of 31\% to $B^*K$
and 28\% to $BK^*$.  In contrast, the $B_s(2P_1)$ has total width of 218~MeV and the $B_s(2^3P_2)$
has a total width of 246~MeV.  Although the $B_s(2P_1)$ and $B_s(2^3P_2)$ overlap with the   
$B_s(2^3P_0)$ and $B_s(2P_1')$, and the $B_s(2P_1)$ has significant BR's to $B^*K$ of 40\% 
and to $BK^*$ of  33\%, and the $B_s(2^3P_2)$ to $BK$ of 12\% and to $B^*K$ of 20\%, it should
be possible to extract a meaningful signal for the $B_s(2^3P_0)$ in the $BK$ final state
and for the $B_s(2P_1')$ in the $B^*K$ final state. 

The $3S$ and $1F$ 
multiplets are very close in mass, at approximately $6300$~MeV, and overlap.  The $B(3^3S_1)$ and $B(3^1S_0)$ have
total widths of 140 and 151~MeV respectively with BR's of BR$(3^3S_1 \to B\pi)=2.2\%$,
BR$(3^3S_1 \to B^*\pi)=3.1\%$ and BR$(3^1S_0 \to B^*\pi)=2.8\%$ so that the signals to observe
these states are rather small. Both states have large BR to $B^*\rho$ of 21\% and 22\% respectively
so these are potentially interesting but also more challenging; as the final state consists 
of three pions it would be necessary to perform a Dalitz plot analysis to reconstruct the
intermediate $\rho$ meson.  The narrow $B(1F)$ states are the $B(1F_3)$ with total width 
106~MeV and BR$(B^*\pi) = 25\%$, and the $B(1^3F_4)$ with total width 110~MeV and
BR$(B^*\pi)=15\%$ and BR$(B\pi)=14\%$.  So in fact the narrow $1F$ states might be more
likely to be observed than the $3S$ states although it will be challenging.  We also
note that these states also have reasonable BR's to 
$B^*\rho$ of 20\% for the $B(1F_3)$
and 47\% for the $B(1^3F_4)$.

The situation for the $B_s(3S)$ states is similar, although perhaps a bit more promising, with 
total widths of 130~MeV for the $B_s(3^3S_1)$ and 121~MeV for the $B_s(3^1S_0)$ with 
relatively small BR's to the simple final states; BR$(B_s(3^3S_1) \to BK)=5.5\%$, 
BR$(B_s(3^3S_1) \to B^*K)=6.9\%$ and BR$(B_s(3^1S_0) \to B^*K)=6.7\%$.   
The narrow $B_s(1F)$ states are the $B_s(1F_3)$ with total width 138~MeV and BR to $B^*K$
of 28\% and the $B_s(1^3F_4)$ with total width of 139~MeV and BR to $BK$ of 18\% and to $B^*K$
of 17\%.  Another prominent decay mode for all these states is $BK^*$.  So although 
there are numerous overlapping resonances, some of these states are narrow enough with a reasonable
signal strength to a simple final state that if one can determine their spin it should be possible
to observe them.

As we move higher in mass, more final states become
kinematically allowed so that BR's to simple states become smaller with some
exceptions, which we note in what follows. We refer the interested reader to Tables \ref{tab:B_12S}-\ref{tab:Bs_1G} for
more details.

The $B(3P_1)$ and $B(3^3P_2)$ are relatively narrow with total widths of 93~MeV and 
88~MeV
respectively with the $B(3P_1)$ having BR$(B^*\pi) = 5.4\%$ and the $B(^3P_2)$ having BR$(B\pi) = 2.9\%$
and BR$(B^*\pi) = 4\%$.  The $B(2D_2)$ has a total width of 104~MeV
with BR$(B^*\pi)=19.6\%$ and the $B(2^3D_3)$ has a total width of 94~MeV with 
BR$(B\pi)=4.3\%$ and BR$(B^*\pi)=8.6\%$.  

The $B_s(4^3S_1)$ has a total width of 104~MeV with BR to $BK$ and $B^*K$ of 2.9\% and 4\%
respectively and the  $B_s(4^1S_0)$ has a total width of 85~MeV with BR to $B^*K$ of 5.3\%.
The $B_s(3P)$ multiplet are all relatively narrow with large BR's to simple final states.
For example, most prominently, the $B_s(3^3P_0)$ has a total width of 51~MeV with BR to $BK$ of 24\% 
and the 
$B_s(3P_1')$ has a total width of 48~MeV with BR to $B^*K$ of 21\%.  The final 
state we note is the $B_s(2^3D_3)$ with total width of 107~MeV and BR to $BK$ and
$B^*K$ of 6.4\% and 13.8\% respectively.

Finally,  we included the strong decay widths for the $1G$ multiplets as the $B(1G_4)$ and $B(1^3G_5)$
are relatively narrow, $\Gamma[B(1G_4)]=96$~MeV and $\Gamma[B(1^3G_5)]= 102$~MeV,
with large BR's to simple final states; $BR[B(1G_4) \to B^*\pi]=24.2\%$, 
$BR[B(1^3G_5) \to B\pi]=12.1\%$ and $BR[B(1^3G_5) \to B^*\pi]=13.2\%$.  These states overlap
with the $3P$ multiplet and are close in mass to the $2F$ states so that it will be necessary
to determine their spins to identify them.  It would be interesting to find these states
as they are high $L$ states analogous to Rydberg states of atomic physics.  Their masses would 
give insights into the confining potential and their splittings on the nature of the spin
dependent potentials.  They would test our understanding of QCD in a region that has not been explored.
In contrast, the $B_s(1G)$ states
are broader, $\sim 200-300$~MeV, so are less likely to be easily found.

The important conclusion we wish to draw from these results is that there are numerous 
excited bottom mesons that are expected to be relatively narrow with significant BR's to 
simple final states.  With sufficient statistics it should be possible to observe many of
these states and improve our knowledge of bottom spectroscopy.  The challenge is that there
is significant overlapping of these states so it will be necessary to perform an angular
distribution analysis to determine the spins of the underlying states to disentangle the
observed final states.

\section{Summary}

The primary purpose of this paper is to calculate the properties of excited $B$ and $B_s$ 
mesons as a guide to help identify newly observed states.  The masses were calculated using
the relativized quark model of Godfrey and Isgur
and an alternative relativistic quark model.  Radiative transition
widths were calculated using a nonrelativistic formalism and wavefunctions from the respective models. Strong
decay widths were calculated with the $^3P_0$ quark creation model coupled with harmonic  oscillator 
wavefunctions that were tuned to reproduce the RMS radius of the relevant hadrons.

Our current experimental knowledge of bottom mesons is rather sparse, having only clearly 
identified the two narrow members of the $B$ and $B_s$ $1P$ multiplets.  Two other excited
states have also been observed and identified with the $B(2S)$ states but they have not been
independently confirmed by a second experiment.

In the near future the LHCb experiment offers the possibility of significantly increasing
our knowledge of excited bottom states.  Numerous $B$ and $B_s$ states with moderate widths 
and with significant branching fractions to simple final states (such as
$B\pi$, $B^*\pi$, $BK$, and $B^*K$) are expected.  However, the spectrum consists of many overlapping states, 
thus measuring the spins of putative signals will be vital for the success of any $B$ spectroscopy program.  
With the high statistics expected in future LHC runs we are optimistic that this can be achieved and that our knowledge 
of the bottom meson spectrum will be significantly expanded.

\acknowledgments

The authors gratefully acknowledge Ted Barnes who was involved in an early iteration of this work.
This research was supported in part by
the Natural Sciences and Engineering Research Council of Canada under Grant No. 121209-2009 SAPIN.



\begin{thebibliography}{99}

 \bibitem{Godfrey:2008nc}
  S.~Godfrey and S.~L.~Olsen,
  ``The Exotic XYZ Charmonium-like Mesons,''
  Ann.\ Rev.\ Nucl.\ Part.\ Sci.\  {\bf 58}, 51 (2008)
  [arXiv:0801.3867 [hep-ph]];
 
  E.~S.~Swanson,
  ``The New heavy mesons: A Status report,''
  Phys.\ Rept.\  {\bf 429}, 243 (2006)
  [hep-ph/0601110].

\bibitem{Eichten:2007qx}
  E.~Eichten, S.~Godfrey, H.~Mahlke and J.~L.~Rosner,
  ``Quarkonia and their transitions,''
  Rev.\ Mod.\ Phys.\  {\bf 80}, 1161 (2008)
  [arXiv:hep-ph/0701208].

\bibitem{Godfrey:2009qe}
  S.~Godfrey,
  ``Topics in Hadron Spectroscopy in 2009,''
  arXiv:0910.3409 [hep-ph].
  
\bibitem{Aaij:2015qla} 
  R.~Aaij {\it et al.} [LHCb Collaboration],
  ``Precise measurements of the properties of the $B_1(5721)^{0,+}$ and $B^\ast_2(5747)^{0,+}$ 
  states and observation of $B^{+,0}\pi^{-,+}$ mass structures,''
  JHEP {\bf 1504}, 024 (2015)
  [arXiv:1502.02638 [hep-ex]].
    
\bibitem{Lang:2015hza} 
  C.~B.~Lang, D.~Mohler, S.~Prelovsek and R.~M.~Woloshyn,
  ``Predicting positive parity B$_s$ mesons from lattice QCD,''
  Phys.\ Lett.\ B {\bf 750}, 17 (2015)
  [arXiv:1501.01646 [hep-lat]].
  
\bibitem{Godfrey:1985xj}
  S.~Godfrey and N.~Isgur,
  ``Mesons In A Relativized Quark Model With Chromodynamics,''
  Phys.\ Rev.\  D {\bf 32}, 189 (1985).

\bibitem{Godfrey:1986wj}
  S.~Godfrey and R.~Kokoski,
  ``The Properties of p Wave Mesons with One Heavy Quark,''
  Phys.\ Rev.\  D {\bf 43}, 1679 (1991).

\bibitem{Godfrey:2004ya}
  S.~Godfrey,
  ``Spectroscopy of $B_c$ mesons in the relativized quark model,''
  Phys.\ Rev.\  D {\bf 70}, 054017 (2004)
  [arXiv:hep-ph/0406228].

\bibitem{Godfrey:2005ww}
  S.~Godfrey,
  ``Properties of the Charmed P-wave Mesons,''
  Phys.\ Rev.\  D {\bf 72}, 054029 (2005)
  [arXiv:hep-ph/0508078].

\bibitem{LS}
For more information on this point see
  O.~Lakhina and E.~S.~Swanson,
  ``Dynamic properties of charmonium,''
  Phys.\ Rev.\ D {\bf 74}, 014012 (2006).
  [hep-ph/0603164].

\bibitem{ls2}
  O.~Lakhina and E.~S.~Swanson,
  ``A Canonical Ds(2317)?,''
  Phys.\ Lett.\ B {\bf 650}, 159 (2007)
  [hep-ph/0608011].

\bibitem{ess}
E.S. Swanson, work in progress.

\bibitem{barnes}
This is discussed more fully in Appendix A of 
T.~Barnes, N.~Black and P.~R.~Page,
``Strong decays of strange quarkonia,''
Phys.\ Rev.\ D {\bf 68}, 054014 (2003)
[arXiv:nucl-th/0208072].

\bibitem{eichten94}
E.J.Eichten and C.Quigg,
``Mesons with beauty and charm: Spectroscopy,''
Phys. Rev. D49, 5845 (1994)
[hep-ph/9402210].
  
\bibitem{nr} 
N. Brambilla, A. Pineda, J. Soto, and A. Vairo, 
``Effective-field theories for quarkonium,"
Rev. Mod. Phys. \textbf{77}, 1423 (2005).

\bibitem{BW}
This form for the spin-dependent interaction was obtained in 
Wilson loop approach by
Eichten and Feinberg,
E. Eichten and F. Feinberg, 
``Spin-dependent forces in quantum chromodynamics,"
Phys. Rev. {\bf D23}, 2724 (1981),
who extended the analysis by Brown and Weisberger, 
L.~S.~Brown and W.~I.~Weisberger, 
``Remarks on the static potential in quantum chromodynamics,"
Phys. Rev. {\bf D20}, 3239
(1979).
Subsequently, Gupta and Radford,
  S.~N.~Gupta and S.~F.~Radford,
  ``Quark Quark And Quark - Anti-Quark Potentials,''
  Phys.\ Rev.\ D {\bf 24}, 2309 (1981),
  S.~N.~Gupta, S.~F.~Radford and W.~W.~Repko,
  ``Quantum Chromodynamic Potential Model For Light And Heavy Quarkonia,''
  Phys.\ Rev.\ D {\bf 28}, 1716 (1983)
(See also
 J.~T.~Pantaleone, S.~H.~H.~Tye and Y.~J.~Ng,
 ``Spin Splittings In Heavy Quarkonia,''
 Phys.\ Rev.\ D {\bf 33}, 777 (1986)),
 performed a one-loop computation of the heavy quark
interaction and showed that a fifth interaction, $V_5$ is present in the case of unequal quark
masses. 

\bibitem{KK}
  Y.~Koma and M.~Koma,
  ``Spin-dependent potentials from lattice QCD,''
  Nucl.\ Phys.\ B {\bf 769}, 79 (2007)
  [hep-lat/0609078].

\bibitem{ss}
For a review of the relationship of QCD to vector and scalar confinement see
  A.~P.~Szczepaniak and E.~S.~Swanson,
  ``On the Dirac structure of confinement,''
  Phys.\ Rev.\ D {\bf 55}, 3987 (1997)
  [hep-ph/9611310].
  

\bibitem{DiPierro:2001dwf} 
  M.~Di Pierro and E.~Eichten,
  ``Excited heavy - light systems and hadronic transitions,''
  Phys.\ Rev.\ D {\bf 64}, 114004 (2001)
  [hep-ph/0104208].

\bibitem{gm}
  A.~Manohar and H.~Georgi,
  ``Chiral Quarks and the Nonrelativistic Quark Model,''
  Nucl.\ Phys.\ B {\bf 234}, 189 (1984);
  J.~L.~Goity and W.~Roberts,
  ``A Relativistic chiral quark model for pseudoscalar emission from heavy mesons,''
  Phys.\ Rev.\ D {\bf 60}, 034001 (1999)
  [hep-ph/9809312].


\bibitem{Close:2005se} 
  F.~E.~Close and E.~S.~Swanson,
  ``Dynamics and decay of heavy-light hadrons,''
  Phys.\ Rev.\ D {\bf 72}, 094004 (2005)
  [hep-ph/0505206].

\bibitem{Matsuki:2007zza} 
  T.~Matsuki, T.~Morii and K.~Sudoh,
  ``New heavy-light mesons Q anti-q,''
  Prog.\ Theor.\ Phys.\  {\bf 117}, 1077 (2007)
  [hep-ph/0605019].


\bibitem{Ebert:2009ua} 
  D.~Ebert, R.~N.~Faustov and V.~O.~Galkin,
  ``Heavy-light meson spectroscopy and Regge trajectories in the relativistic quark model,''
  Eur.\ Phys.\ J.\ C {\bf 66}, 197 (2010)
  [arXiv:0910.5612 [hep-ph]].

\bibitem{Devlani:2012zz} 
  N.~Devlani and A.~K.~Rai,
  ``Spectroscopy and decay properties of B and B/s mesons,''
  Eur.\ Phys.\ J.\ A {\bf 48}, 104 (2012).

\bibitem{Xiao:2014ura} 
  L.~Y.~Xiao and X.~H.~Zhong,
  ``Strong decays of higher excited heavy-light mesons in a chiral quark model,''
  Phys.\ Rev.\ D {\bf 90}, no. 7, 074029 (2014)
  [arXiv:1407.7408 [hep-ph]].

\bibitem{Sun:2014wea} 
  Y.~Sun, Q.~T.~Song, D.~Y.~Chen, X.~Liu and S.~L.~Zhu,
  ``Higher bottom and bottom-strange mesons,''
  Phys.\ Rev.\ D {\bf 89}, no. 5, 054026 (2014)
  [arXiv:1401.1595 [hep-ph]].

\bibitem{Ferretti:2015rsa} 
  J.~Ferretti and E.~Santopinto,
  ``Open-flavor strong decays of open-charm and open-bottom mesons in the $^3P_0$ pair-creation model,''
  arXiv:1506.04415 [hep-ph].

\bibitem{Liu:2015lka} 
  J.~B.~Liu and M.~Z.~Yang,
  ``Spectrum of Higher excitations of $B$ and $D$ mesons in the relativistic potential model,''
  Phys.\ Rev.\ D {\bf 91}, no. 9, 094004 (2015)
  [arXiv:1501.04266 [hep-ph]].


\bibitem{Liu:2016efm} 
  J.~B.~Liu and C.~D.~Lu,
  ``Spectra of heavy-light mesons in a relativistic model,''
  arXiv:1605.05550 [hep-ph].

\bibitem{Lu:2016bbk} 
  Q.~F.~L\"{u}, T.~T.~Pan, Y.~Y.~Wang, E.~Wang and D.~M.~Li,
  ``Excited bottom and bottom-strange mesons in the quark model,''
  arXiv:1607.02812 [hep-ph].



\bibitem{Kwo88}
See for example 
W.Kwong and J.L.Rosner,
``D Wave Quarkonium Levels Of The Upsilon Family,''
Phys. Rev. D38, 279 (1988).

\bibitem{Sie37}
A.J.Siegert,
``Note On The Interaction Between Nuclei And Electromagnetic Radiation,''
Phys. Rev. 52, 787 (1937).

\bibitem{McC83}
R.McClary and N.Byers,
``Relativistic Effects In Heavy Quarkonium Spectroscopy,''
Phys. Rev. D28, 1692 (1983).

\bibitem{Mox83}
P.Moxhay and J.L.Rosner,
``Relativistic Corrections In Quarkonium,''
Phys. Rev. D28, 1132 (1983).

\bibitem{JDJ} J. D. Jackson, ``Lecture on the New Particles,'' in {\it
Proceedings of the Summer Institute on Particle Physics, August 2--13, 1976},
edited by M. C. Zipf, Stanford Linear Accelerator Center Report SLAC-198,
November 1977, p.~147.

\bibitem{Nov78}
V.A.~Novikov, L.B.~Okun, M.A.~Shifman, A.I.~Vainshtein, M.B.~Voloshin, 
and V.I.~Zakharov, 
``Charmonium and gluons,"
Phys. Rept. C41, 1 (1978).

\bibitem{m1}
Relativistic effects in M1 transitions are discussed in:  
J.~S.~Kang and J.~Sucher,
``Radiative M1 Transitions Of The Narrow Resonances,''
Phys.\ Rev.\ D {\bf 18}, 2698 (1978);
H.~Grotch and K.~J.~Sebastian,
``Magnetic Dipole Transitions Of Narrow Resonances,''
Phys.\ Rev.\ D {\bf 25} 2944 (1982) ;
V.~Zambetakis and N.~Byers,
``Magnetic Dipole Transitions In Quarkonia,''
Phys.\ Rev.\ D {\bf 28}, 2908 (1983);
H.~Grotch, D.~A.~Owen and K.~J.~Sebastian,
 ``Relativistic Corrections To Radiative Transitions And Spectra Of Quarkonia,''
Phys.\ Rev.\ D {\bf 30}, 1924 (1984);
X.~Zhang, K.~J.~Sebastian and H.~Grotch,
``M1 Decay Rates Of Heavy Quarkonia With A Nonsingular Potential,''
Phys.\ Rev.\ D {\bf 44}, 1606 (1991).

\bibitem{ebert03}
D.Ebert, R.N.Faustov and V.O.Galkin,
``Properties of heavy quarkonia and B/c mesons in the relativistic quark model,''
Phys. Rev. D67, 014027 (2003)
[hep-ph/0210381].

\bibitem{Micu:1968mk} 
  L.~Micu,
  ``Decay rates of meson resonances in a quark model,''
  Nucl.\ Phys.\ B {\bf 10}, 521 (1969).
  
\bibitem{Le Yaouanc:1972ae} 
  A.~Le Yaouanc, L.~Oliver, O.~Pene and J.~C.~Raynal,
  ``Naive quark pair creation model of strong interaction vertices,''
  Phys.\ Rev.\ D {\bf 8}, 2223 (1973).
  
\bibitem{Ackleh:1996yt} 
  E.~S.~Ackleh, T.~Barnes and E.~S.~Swanson,
  ``On the mechanism of open-flavor strong decays,''
  Phys.\ Rev.\ D {\bf 54}, 6811 (1996)
  [hep-ph/9604355].

\bibitem{Blundell:1995ev} 
  H.~G.~Blundell and S.~Godfrey,
  ``The Xi (2220) revisited: Strong decays of the $1^3F_2$ and $1^3F_4$ $s\bar{s}$ mesons,''
  Phys.\ Rev.\ D {\bf 53}, 3700 (1996)
  [hep-ph/9508264].

  \bibitem{Barnes:2005pb} 
  T.~Barnes, S.~Godfrey and E.~S.~Swanson,
  ``Higher charmonia,''
  Phys.\ Rev.\ D {\bf 72}, 054026 (2005)
  [hep-ph/0505002].

\bibitem{Geiger:1994kr} 
  P.~Geiger and E.~S.~Swanson,
  ``Distinguishing among strong decay models,''
  Phys.\ Rev.\ D {\bf 50}, 6855 (1994)
  [hep-ph/9405238].

\bibitem{Olive:1900zz} 
  K.~A.~Olive {\it et al.}  [Particle Data Group Collaboration],
  ``Review of Particle Physics (RPP),''
  Chin.\ Phys.\ C {\bf 38}(9), 090001 (2014).

   \bibitem{Godfrey:2015} 
  S.~Godfrey and K.~Moats,
  ``Bottomonium Mesons and Strategies for their Observation,''
  Phys.\ Rev.\ D {\bf 92},  054034 (2015)
  [arXiv:1507.00024 [hep-ph]].
  

\bibitem{Blundell:1996as} 
  H.~G.~Blundell,
  ``Meson properties in the quark model: A look at some outstanding problems,''
  hep-ph/9608473.

 \bibitem{Godfrey:2014fga} 
  S.~Godfrey and K.~Moats,
  ``$D_{sJ}^*(2860)$ mesons as excited $D$-wave $c\bar{s}$ states,''
  Phys.\ Rev.\ D {\bf 90}, 117501 (2014); {\bf 92}, 119903(E) (2015)
  [arXiv:1409.0874 [hep-ph]].

\bibitem{Godfrey:2015dva} For a recent overview of this topic see for example
  S.~Godfrey and K.~Moats,
  ``Properties of Excited Charm and Charm-Strange Mesons,''
  Phys.\ Rev.\ D {\bf 93}, no. 3, 034035 (2016)
  [arXiv:1510.08305 [hep-ph]].

\end{thebibliography}
\end{document}